\documentclass[conference]{IEEEtran}
\IEEEoverridecommandlockouts
\usepackage{cite}
\usepackage{amsmath,amssymb,amsfonts}
\usepackage{graphicx}
\usepackage{textcomp}
\usepackage{xcolor}
\usepackage{subfigure}
\usepackage{multicol}
\usepackage{multirow}
\usepackage[caption=false,font=normalsize,labelfont=sf,textfont=sf]{subfig}
\usepackage{balance}
\usepackage{algorithm}
\usepackage{algpseudocode}

\def\BibTeX{{\rm B\kern-.05em{\sc i\kern-.025em b}\kern-.08em
    T\kern-.1667em\lower.7ex\hbox{E}\kern-.125emX}}
\usepackage{booktabs}

\begin{document}

\title{Text-Guided Diffusion Model-based Generative Communication for Wireless Image Transmission}

\author{Shengkang Chen, Tong Wu, Zhiyong Chen, Feng Yang, Meixia Tao, \emph{IEEE Fellow}, Wenjun Zhang, \emph{IEEE Fellow}\\
        \thanks{The authors are with the Cooperative Medianet Innovation Center and the Department of Electronic Engineering, Shanghai Jiao Tong University, Shanghai 200240, China (e-mail: \{chen20,wu\_tong,zhiyongchen,yangfeng, mxtao, zhangwenjun\}@sjtu.edu.cn). }}

\maketitle

\begin{abstract}
Reliable image transmission over wireless channels is particularly challenging at extremely low transmission rates, where conventional compression and channel coding schemes fail to preserve adequate visual quality. To address this issue, we propose a generative communication framework based on diffusion models, which integrates joint source–channel coding (JSCC) with semantic-guided reconstruction leveraging a pre-trained generative model. Unlike conventional architectures that aim to recover exact pixel values of the original image, the proposed method focuses on preserving and reconstructing semantically meaningful visual content under severely constrained rates, ensuring perceptual plausibility and faithfulness to the scene intent. Specifically, the transmitter encodes the source image via JSCC and jointly transmits it with a textual prompt over the wireless channel. At the receiver, the corrupted low-rate representation is fused with the prompt and reconstructed through a Stable Diffusion model with ControlNet, enabling high-quality visual recovery. Leveraging both generative priors and semantic guidance, the proposed framework produces perceptually convincing images even under extreme bandwidth limitations. Experimental results demonstrate that the proposed method consistently outperforms conventional coding-based schemes and deep learning baselines, achieving superior perceptual quality and robustness across various channel conditions.
\end{abstract}

\begin{IEEEkeywords}
Pretrained diffusion model, image transmission, low transmission rates.
\end{IEEEkeywords}

\section{Introduction}
Wireless communication demands have grown substantially in recent decades, driven by data-intensive applications such as remote sensing, autonomous driving, and multimedia services. Traditional communication technologies often struggle to meet these ever-increasing requirements, particularly in extreme scenarios such as underwater channels and satellite links, where bandwidth is severely constrained and channel impairments are significant. These limitations highlight the need for novel paradigms capable of maintaining semantic fidelity and perceptual quality under ultra-low bitrate constraints. Semantic communication has thus emerged as a promising direction, aiming to enhance efficiency by conveying the underlying meaning of information rather than transmitting raw data \cite{Zhang2025,9955525}.

Among the various approaches to realizing semantic communication, deep learning-based joint source–channel coding (JSCC) has attracted considerable attention \cite{10183794,10012892}. Unlike conventional image transmission pipelines that separate compression and error correction, JSCC directly maps source data into channel symbols through a compact latent representation. Early work such as DeepJSCC \cite{9954153} demonstrated the potential of neural JSCC, outperforming traditional separate coding schemes that combine advanced image codecs like Better Portable Graphics (BPG) with low-density parity-check (LDPC) codes. Building on this foundation, subsequent studies have extended JSCC with attention mechanisms for channel adaptivity (ADJSCC \cite{9438648}), vision transformers and entropy models for semantic compression (NTSCC \cite{9791398}), hierarchical transformer backbones for improved robustness (SwinJSCC \cite{10094735}), and state space models for parameter-efficient architectures (MambaJSCC \cite{10901192}).

Despite these advances, the performance of deep JSCC remains fundamentally constrained at extremely low bitrates. In such cases, heavy compression of the latent representation leads to blurry reconstructions that lack fine details and structural fidelity, causing the outputs to deviate from the natural image distribution. Conventional decoders trained with pixel-wise or perceptual losses often yield distorted and semantically ambiguous results, unable to capture the rich statistical patterns of real-world visual data. Generative models, on the other hand, excel at modeling complex, high-dimensional image distributions and provide strong priors that encode realistic textures, structures, and semantic coherence. Leveraging these generative priors offers a promising pathway to refine coarse JSCC reconstructions into visually plausible and perceptually faithful images, motivating the integration of generative models into JSCC frameworks.


Recently, diffusion models have demonstrated remarkable generative capabilities in both image synthesis and restoration tasks \cite{ho2020denoising, song2020denoising}. Large-scale pre-trained models such as Stable Diffusion \cite{Rombach_2022_CVPR} achieve impressive results in unconditional and conditional generation. By leveraging powerful generative priors learned from massive image–text datasets, these models can produce semantically consistent and visually realistic outputs conditioned on textual prompts or auxiliary visual inputs. Such strengths make diffusion models highly appealing for wireless image transmission, where they hold the potential to enhance perceptual quality and preserve semantic fidelity beyond the capabilities of conventional JSCC decoders.

Although diffusion models are typically trained under the assumption of Gaussian noise corruption, recent studies have shown their effectiveness in addressing more complex, non-Gaussian distortions. Soft Diffusion \cite{daras2022soft} extends score matching to a broader family of linear corruptions, including blurring and masking, thereby enabling robust restoration under structured degradations. Building on this idea, \cite{stevens2023removing} demonstrated that diffusion models can be adapted to mitigate structured noise through joint posterior sampling. Even in the case of nonlinear and non-invertible distortions such as severe JPEG artifacts, CODiff \cite{guo2025compression} has been shown to successfully reconstruct original images by incorporating compression-aware priors into a one-step diffusion framework. These advances suggest that, although JSCC reconstructions deviate from the standard Gaussian corruption assumption, diffusion models can still be effectively exploited to refine and denoise such outputs, provided that the coarse semantic structure is preserved.

Inspired by these successes, researchers have recently explored the application of diffusion models in wireless communication. For example, CDDM \cite{10480348} employs a diffusion model to denoise the received signal at the decoder, while \cite{10620904} applies diffusion-based post-processing modules to refine degraded images at the receiver. DiffCom \cite{10960324} further incorporates diffusion-based denoising strategies for image transmission over wireless channels. However, since DiffCom does not exploit explicit semantic conditions, its reconstructions may appear perceptually plausible yet semantically inconsistent under ultra-low bitrate constraints. More recently, SGD-JSCC \cite{11099589} introduced semantics-guided diffusion for JSCC, where edge maps are used as semantic guidance during diffusion sampling on JSCC-generated latent features. Although this approach improves performance in low-SNR channels, the reliance on edge maps as conditional inputs inevitably discards color and fine-grained texture information, and its evaluation is limited to slow fading channels.

Beyond visual priors, advances in multimodal large language models (MLLMs) provide strong semantic understanding and text generation capabilities. Trained on massive multimodal datasets, these models can generate accurate and contextually rich textual descriptions of images, which in turn serve as powerful prompts for guiding generative models. In this paper, we leverage an MLLM to automatically produce descriptive prompts for transmitted images, thereby eliminating the need for manual annotations. This integration ensures that the diffusion model is guided by reliable semantic cues, enabling reconstructions that are not only perceptually realistic but also semantically faithful.

In this paper, we applies diffusion sampling in the pixel space, using the JSCC-decoded blurry image together with textual prompts as conditional inputs to a pre-trained Stable Diffusion model equipped with ControlNet. This design enables the recovery of high-quality, semantically consistent images under ultra-low bitrate conditions without retraining the diffusion model. Although the JSCC-decoded image lacks fine details, it still preserves coarse spatial structures, color distributions, and global layout information that are critical for faithful reconstruction. By feeding this image into the diffusion model through the ControlNet conditioning mechanism, our framework effectively maintains the geometric and chromatic consistency of the original scene. This structural guidance complements the high-level semantic constraints provided by textual prompts, forming a dual-path conditioning strategy that jointly enforces both visual plausibility and semantic accuracy.

Textual prompts serve as explicit semantic anchors that guide the diffusion process toward reconstructions faithful to the original scene semantics. Unlike implicit reconstructions from JSCC alone, which may drift semantically due to ambiguous latent representations, text-based conditions provide structured, human-interpretable constraints that help preserve critical objects and attributes during the detail hallucination process. This capability is particularly crucial under ultra-low bitrate settings, where most visual information is lost and the risk of semantic distortion is significant.

The main contributions of this paper are summarized as follows:

\begin{itemize}
    \item  We propose a generative communication framework based on diffusion models. Unlike prior diffusion-based communication methods, the proposed approach integrates structural guidance from the JSCC-decoded coarse latent with explicit semantic control via textual prompts generated by an MLLM at the receiver. This dual-conditioning strategy enables high-quality and semantically faithful image reconstruction under ultra-low bitrate constraints, without retraining the diffusion model.
    \item  We develop a noise-aware initialization strategy to address the mismatch between distortions introduced by encoding and channel noise and the Gaussian noise prior assumed by diffusion models. By injecting controlled Gaussian noise into the JSCC output, an appropriate starting point for the reverse diffusion process is established, which significantly accelerates convergence. As a result, high-quality refinement is achieved in only $5$ denoising steps, making the framework highly efficient and suitable for low-latency communication scenarios.
    
    \item We conduct extensive experiments demonstrating that the proposed method consistently surpasses conventional JSCC approaches, achieving superior perceptual quality and semantic fidelity under extremely low bitrate conditions.
\end{itemize}

The remainder of this paper is organized as follows. Section~II presents the proposed system model, including the JSCC-based wireless transmission framework and the diffusion-based refinement process. Section~III reports experimental results and analysis, including comparisons with state-of-the-art methods, ablation studies, and visual evaluations. Finally, Section~IV concludes the paper.

\section{System Model}



\subsection{Overall framework}

\begin{figure*}[t]
    \centering
    \includegraphics[width=0.9\linewidth]{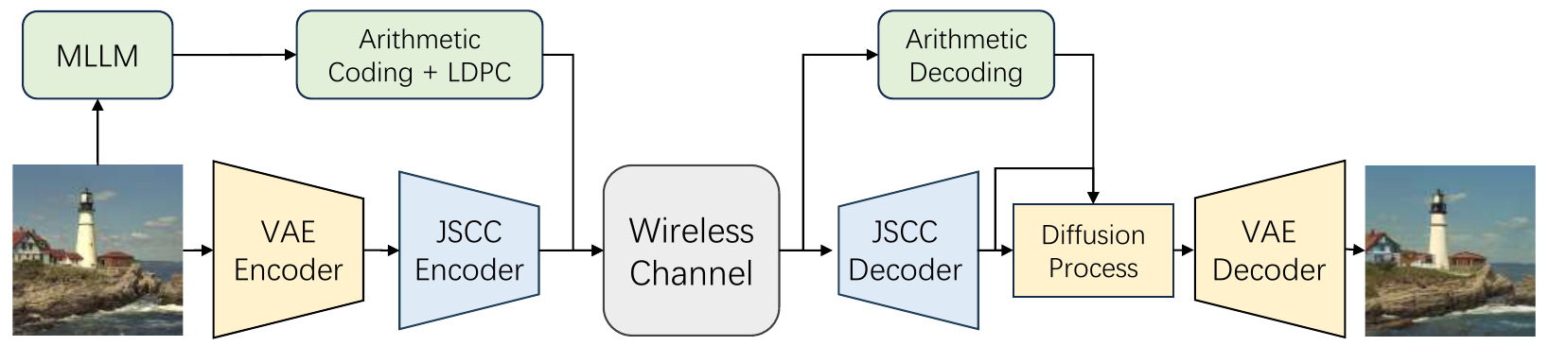}
    \caption{Overview of the proposed system model.}
    \label{fig:frame}
\end{figure*}

The overall architecture of the proposed system is shown in Fig.~\ref{fig:frame}. At the transmitter, the source image $\mathbf s \in \mathbb{R}^{3 \times H \times W}$ is transformed into a compact latent representation through the pretrained variational autoencoder (VAE) component of the Stable Diffusion model. The resulting vector $\mathbf z_c \in \mathbb{R}^{2k'}$ is subsequently processed by the JSCC encoder to produce a real-valued transmit vector $\mathbf x \in \mathbb{R}^{2k}$. Finally, the real-valued vector $\mathbf x$ is converted into a complex-valued signal $\mathbf x_c \in \mathbb{C}^k$ by assigning its first $k$ elements to the real part and the remaining $k$ elements to the imaginary part.

In parallel, a multimodal large language model generates a concise textual prompt $o$ that summarizes the high-level semantic attributes of the input image, such as objects, actions, and contextual information. The prompt $o$ is then compressed using arithmetic coding, protected with LDPC channel coding, and transmitted over an orthogonal channel with $k_o$ additional channel uses.

We consider both additive white Gaussian noise (AWGN) and Rayleigh fading channels. The received signal can be expressed as
\begin{equation}
    \mathbf{y}_c = \mathbf{h} \odot \mathbf{x}_c + \mathbf{n},
\end{equation}
where $\mathbf{h} \in \mathbb{C}^k$ denotes the channel gain vector, $\mathbf{n} \sim \mathcal{CN}(0,\sigma^2\mathbf{I})$ is the noise vector, and $\odot$ represents element-wise multiplication. For the AWGN channel, $\mathbf{h} = \mathbf{1}$, whereas for the Rayleigh fading channel, the elements of $\mathbf{h}$ are i.i.d. $\mathcal{CN}(0,1)$.

At the receiver, the received complex signal is first equalized using a minimum mean square error (MMSE) equalizer and then decoded by the JSCC decoder to obtain a coarse reconstruction of the latent representation $\hat{\mathbf{z}}_c$. 
 Simultaneously, the received text bitstream is decoded to recover the prompt $o$.

Finally, in the subsequent diffusion-based refinement stage, the coarse latent $\hat{\mathbf{z}}$ provides structural guidance, while the prompt $o$ supplies semantic guidance to the diffusion model. To accelerate sampling, $\hat{\mathbf{z}}_c$ is perturbed with controlled Gaussian noise to generate an initial latent $z_{N_s}$ corresponding to a selected intermediate diffusion step $N_s$, where $N_s$ is a hyperparameter chosen based on the channel bandwidth ratio (CBR). This warm-start strategy ensures that the diffusion process begins from a semantically consistent state, enabling high-fidelity image recovery within only $5$ denoising steps.

\subsection{JSCC-based Wireless Image Transmission}

For JSCC, we adopt the MambaJSCC scheme \cite{10901192} in this paper, which employs a Visual State Space Model with Channel Adaptation (VSSM-CA) as its backbone. VSSM-CA integrates 2D image features into a state-space representation, enabling efficient encoding with linear computational complexity. It further incorporates channel state information (CSI) via a lightweight integration mechanism, allowing robust adaptation to varying channel conditions. Compared to other JSCC methods, such as SwinJSCC, MambaJSCC achieves higher PSNR performance while maintaining substantially lower computational complexity and inference latency. These advantages have been validated through extensive experiments on benchmark datasets.


\section{The proposed strategy}
\subsection{Latent JSCC-based Transmission}
Instead of transmitting the reconstructed image $\hat{\mathbf{s}}$ in the pixel domain, we explore a latent-domain design in which JSCC is performed directly on the latent space of the pretrained VAE encoder in Stable Diffusion. Specifically, the source image $\mathbf{s}$ is first mapped into a latent representation $z = \mathcal{E}(\mathbf{s})$ by the VAE encoder $\mathcal{E}$ of Stable Diffusion. This latent $z$ is then transmitted over the wireless channel via a JSCC encoder–decoder pair.
The JSCC encoder $f_{\theta}(\cdot)$ maps $z$ to a semantic feature vector $\mathbf{x} \in \mathbb{R}^{2k}$:
\begin{equation}
    \mathbf{x} = f_{\theta}(z),
\end{equation}
where $\theta$ denotes the trainable parameters of the JSCC encoder. The vector $\mathbf{x}$ is then converted into a complex-valued channel input $\mathbf{x}_c \in \mathbb{C}^k$. After transmitted through the wireless channel, the received symbols $\mathbf{y}_c$ are equalized and reshaped to $\mathbf{y}$, which is subsequently decoded by the JSCC decoder $g_{\phi}(\cdot)$ to produce a coarse image reconstruction:
\begin{equation}
    z_c = g_{\phi}(\mathbf{y}).
\end{equation}
where $\phi$ denotes the trainable parameters of the JSCC encoder.
After channel transmission and decoding, the recovered latent $z_c$ is obtained, which serves as the starting point for the diffusion process during subsequent denoising and refinement.

This latent-domain design offers several advantages. First, because $z$ is already a compact representation of $\mathbf{s}$, the computational burden is reduced, enabling higher efficiency at ultra-low bitrates. Second, the latent representation is naturally aligned with the diffusion model, which operates natively in the same latent space. Compared to transmitting $\hat{\mathbf{s}}$ in the pixel domain, latent JSCC avoids introducing artifacts that could confuse the diffusion refinement and provides a cleaner starting point for semantic-conditioned reconstruction.

An additional advantage of the latent-domain design is its scalability to higher resolutions. Unlike pixel-domain JSCC, which is typically resolution-specific, the latent-based model trained on 256$\times$256 images can be directly applied to 512$\times$512 inputs. Under the same channel bandwidth ratio (CBR), the latent representation of higher-resolution images results in a larger number of transmitted symbols, allowing more structural information to be preserved during transmission. Consequently, the reconstructed high-resolution images exhibit significantly improved fidelity and richer details.

\subsection{Semantic Prompt Generation via MLLM}

\begin{figure}
    \centering
    \includegraphics[width=1.0\linewidth]{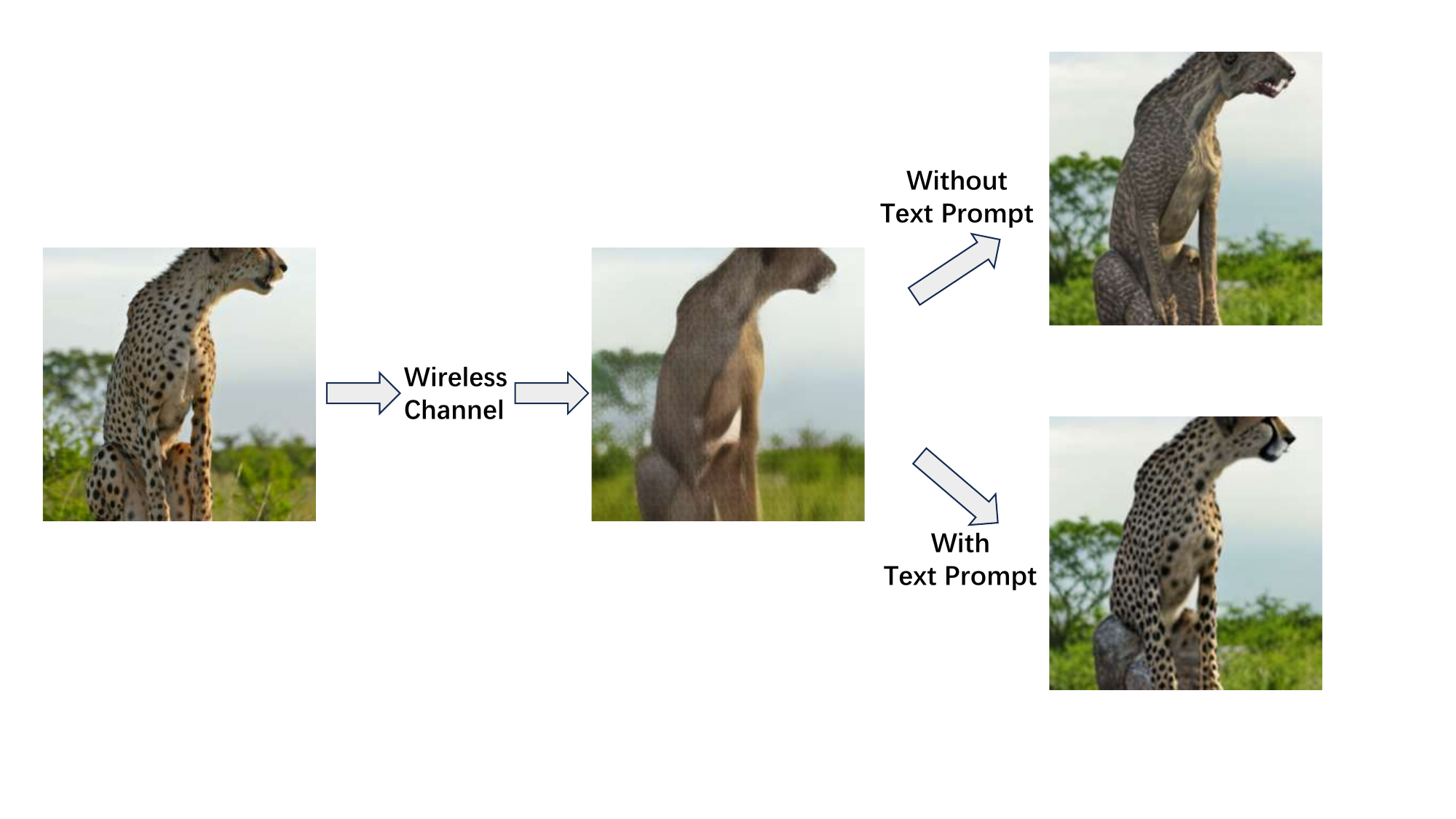}
    \caption{The impact of semantic prompts on diffusion-based refinement.}
    \label{fig:prompt}
\end{figure}

Although the coarse reconstruction $z_c$ obtained from JSCC preserves the coarse color and structural information of the original image $\mathbf{s}$, it is often insufficient for faithful recovery under ultra-low bitrate conditions. In particular, the loss of fine details frequently leads to blurred textures and ambiguous object boundaries, making it difficult to accurately infer the underlying semantics from $\hat{\mathbf{s}}$ alone. For example, as shown in Fig.~\ref{fig:prompt}, the source image clearly depicts a cheetah. However, after transmission through the wireless channel, the reconstructed image becomes severely blurred, making it challenging to recognize the animal. Moreover, applying the diffusion model directly to this degraded latent without additional semantic guidance often fails to preserve the identity of the cheetah. This motivates the incorporation of an additional modality that provides explicit semantic guidance.

To address this, we leverage the MLLM to extract high-level textual descriptions from the source image $\mathbf{s}$. Specifically, we employ GPT-4o to generate a concise prompt $o$ that captures key semantic attributes, including objects, actions, and contextual information present in the scene. Unlike the coarse reconstruction $z_c$, which may preserve only global structures while losing fine-grained details due to channel-induced distortions, the prompt $o$ focuses on detail-oriented semantic attributes that are otherwise unrecoverable from $\hat{\mathbf{s}}$. Ensuring the semantic accuracy of $o$ is crucial; incorrect or overly generic prompts may mislead the diffusion model, resulting in refinements that deviate from the source content.

Since errors in the transmitted prompt could drastically alter the semantic meaning of the image, reliable delivery is necessary. To ensure robustness, the prompt $o$ is first compressed using arithmetic coding to remove redundancy and then protected with LDPC channel coding before transmission. The encoded bitstream requires $k_o$ additional channel uses, which are allocated orthogonally to the image transmission channel. At the receiver, the decoded prompt $\hat{o}$ is combined with the coarse reconstruction $\hat{\mathbf{s}}$ to provide dual-modal guidance for the subsequent diffusion-based refinement stage.

\subsection{Diffusion-Based Image Reconstruction}
We employ a conditional diffusion model to generate the final high-fidelity reconstruction $\tilde{\mathbf{s}}$. The latent representation $z_c$ is obtained directly from the JSCC decoder output. For the diffusion-based refinement process, a noisy version of $z_c$ serves as the starting point, while the original $z_c$ is also used as a conditioning signal.

In standard diffusion models, the forward process gradually corrupts a clean latent feature $z_0$ by adding Gaussian noise at each time step $t$, controlled by a variance schedule $\{\beta_t\}_{t=1}^T$, where $T$ denotes the total number of steps. Formally, the latent variable at step $t$ is given by
\begin{equation}
    z_t = \sqrt{\bar{\alpha}_t} z_0 + \sqrt{1-\bar{\alpha}_t}\epsilon, \quad \epsilon \sim \mathcal{N}(0,\mathbf{I}),
\end{equation}
where $\alpha_t = 1-\beta_t$ and $\bar{\alpha}_t = \prod_{i=1}^t \alpha_i$. As $t$ increases, $z_t$ gradually approaches a pure Gaussian distribution.

In contrast, our goal is faithful reconstruction rather than genrative diversity. Initializing the reverse diffusion from pure Gaussian noise would discard the valuable structural priors embedded in the coarse reconstruction $\hat{\mathbf{s}}$ and introduce unnecessary randomness. To address this, we design a \textbf{warm-start strategy}, in which the reverse process is initialized from a noisy version of the JSCC-reconstructed latent. The initial state is then defined as
\begin{equation}
    z_{N_s} = \sqrt{\bar{\alpha}_{N_s}} z_c + \sqrt{1-\bar{\alpha}_{N_s}}\epsilon, \quad N_s < T,
\end{equation}
where $N_s$ represents the intermediate diffusion step which is chosen according to the CBR.

This warm-start strategy enables the reverse process to begin from a \textbf{semantically consistent latent representation}, thereby accelerating convergence and producing high-quality reconstructions within only a few denoising steps, as also observed in prior work~\cite{yue2023resshift,yang2024pixel}. In our method, this approach produces high-quality reconstructions in as few as 5 denoising steps. Empirically, we observe that increasing $N_s$ generally enhances perceptual quality by providing a stronger denoising trajectory. However, beyond a certain threshold, additional increases in $N_s$ yield diminishing returns, with the reconstruction quality remaining nearly unchanged. This indicates that a moderate choice of $N_s$ effectively balances visual fidelity and computational efficiency. 
 
Accordingly, we further interpret the difference between the clean latent $z_0$ and the coarse latent $z_c$ as a structured residual noise component. Formally, we define this residual as $\epsilon_{\text{res}} = z_c - z_0$, and reformulate the forward process as
\begin{equation}
q(z_t|z_0,\epsilon_{res})\sim \mathcal{N}(\sqrt{\bar{\alpha}_t} z_0 + \sqrt{1-\bar{\alpha}_t}\gamma_t \epsilon_{\text{res}} ,(1-\bar{\alpha}_t)\mathbf{I})
\end{equation}
where $\epsilon_t \sim \mathcal{N}(0,I)$ denotes standard Gaussian noise and $\gamma_t$ is a weighting factor that controls the contribution of the residual noise. The latent variable at step $t$ can be sampled as:
\begin{equation}
    z_t = \sqrt{\bar{\alpha}_t} z_0 + \sqrt{1-\bar{\alpha}_t}\left(\gamma_t \epsilon_{\text{res}} + \epsilon_t\right), 
    \quad t \in \{1,\dots,N_s\}.
    \label{comb1}
\end{equation}

From (\ref{comb1}), for $n < N_s$, the intermediate latent variable $z_t$ can be expressed as  
\begin{equation}
\begin{aligned}
    z_t &= \sqrt{\bar{\alpha}_t}z_0 + \sqrt{1-\bar{\alpha}_t}\left(\gamma_t \epsilon_{\text{res}} + \epsilon_t\right)\\
    &= \sqrt{\bar{\alpha}_t}z_0 + \sqrt{1-\bar{\alpha}_t}\gamma_t (z_c-z_0) + \sqrt{1-\bar{\alpha}_t}\epsilon_t\\
    &= \big(\sqrt{\bar{\alpha}_t}-\sqrt{1-\bar{\alpha}_t}\gamma_t\big)z_0 
       + \sqrt{1-\bar{\alpha}_t}\gamma_t z_c 
       + \sqrt{1-\bar{\alpha}_t}\epsilon_t .
\end{aligned}
\end{equation}

Hence, $z_0$ can be recovered as  
\begin{equation}
z_0=\frac{1}{\sqrt{\bar{\alpha}_t}-\sqrt{1-\bar{\alpha}_t}\gamma_t}\Big[z_t-\sqrt{1-\bar{\alpha}_t}(\gamma_tz_c+\epsilon_t )\Big], 
\quad n< N_s.
\label{eq:pred_z0}
\end{equation}

The resulting forward process is non-Markovian, as each latent state depends jointly on both the clean latent $z_0$ and the coarse latent $z_c$. Specifically, $z_{t-1}$ and $z_t$ are conditionally independent given $z_0$ and $z_c$, since $z_{t-1}$ is determined by $z_0$, $z_c$, and an independent noise term $\epsilon_{t-1}$ that does not depend on $z_t$. Accordingly, the conditional distribution can be written as
$
q(z_{t-1}\mid z_t, z_0, z_c) = q(z_{t-1}\mid z_0, z_c) 
\sim\mathcal{N}(\sqrt{\bar{\alpha}_{t-1}} z_0 + \sqrt{1-\bar{\alpha}_{t-1}}\gamma_t \epsilon_{\text{res}}, (1-\bar{\alpha}_{t-1})I).
$

To facilitate practical sampling, we parameterize this conditional distribution in a linear form as
\begin{equation}
q(z_{t-1} \mid z_t, z_0, z_c) \sim \mathcal{N}(a_t z_t + b_t z_0, \sigma_t^2 \mathbf{I}),
\label{comb2}
\end{equation}
where the coefficients $a_t$, $b_t$, and the variance $\sigma_t^2$ are selected such that this formulation remains equivalent to the forward process defined above.

\begin{algorithm}[t]
\caption{Sampling Algorithm} 
\label{alg:1} 
\begin{algorithmic}[1] 
\Statex\hspace*{-15pt}\textbf{Input:} Latent $\hat z_c$, 
         textual prompt $o$, diffusion model $\epsilon_\theta$, 
         number of steps $N$, warm-start step $N_s$ (hyperparameter), guidance scale $\omega$
\Statex \hspace*{-15pt}\textbf{Output:} Reconstructed high-fidelity image $\tilde{\mathbf{s}}$ 
\State $z_{N_s} = \sqrt{\bar{\alpha}_{N_s}} z_c + \sqrt{1-\bar{\alpha}_{N_s}} \epsilon, \quad \epsilon \sim \mathcal{N}(0,I)$.
\State $t=N_s$,
\While {$t>0$}
\State Calculate $a_t, b_t, \gamma_t$ according to the Eq.~(\ref{eq:a}), (\ref{eq:b}) and (\ref{eq:c}).
\State $\epsilon_{cond}=\epsilon_\theta(z_t, z_c, o, t)$,
\State $\epsilon_{uncond}=\epsilon_\theta(z_t, t)$,
\State $\hat\epsilon_t=\epsilon_{uncond}+\omega(\epsilon_{cond}-\epsilon_{uncond})$,
\State $\hat{z}_0 =$ $\dfrac{1}{\sqrt{\bar{\alpha}_t}-\sqrt{1-\bar{\alpha}_t}\gamma_t}\Big[z_t-\sqrt{1-\bar{\alpha}_t}(\gamma_tz_c+ \hat\epsilon_t)\Big]$,
\State $z_{t-\frac{N_s}{N}} = a_t z_t + b_t \hat{z}_0$,
\State $t= t-\frac{N_s}{N}$,
\EndWhile
\State $\tilde{\mathbf{s}} \gets \mathcal{D}(z_0)$,
\State \Return $\tilde{\mathbf{s}}$
\end{algorithmic}
\end{algorithm}

Following the DDIM framework~\cite{song2020denoising}, we define the variance term as
\begin{equation}
    \sigma_t = \eta \sqrt{\frac{1-\bar{\alpha}_{t-1}}{1-\bar{\alpha}_t}} \sqrt{1 - \frac{\bar{\alpha}_t}{\bar{\alpha}_{t-1}}},
\end{equation}
where $\eta$ is a hyperparameter controlling the stochasticity of the sampling process. Setting $\eta = 0$ yielding a deterministic sampling trajectory, while $\eta = 1$ corresponds to a fully stochastic one. In this paper, we always set $\eta = 0$ to ensure stable and consistent reconstructions.

By combining (\ref{comb1}) and~(\ref{comb2}), the coefficients $a_t$, $b_t$, and the residual weighting factor $\gamma_t$ can be derived from the following system of equations:
\begin{equation}
\left\{
\begin{aligned}
    &a_t \sqrt{\bar{\alpha}_t} + b_t = \sqrt{\bar{\alpha}_{t-1}},\\
    &a_t^2 (1-\bar{\alpha}_t) + \sigma_t^2 = 1 - \bar{\alpha}_{t-1},\\
    &\sqrt{1-\bar{\alpha}_{t-1}} \gamma_{t-1} = a_t \gamma_t \sqrt{1-\bar{\alpha}_t},\\
    &\sigma_t=0.
\end{aligned}
\right.
\end{equation}

Solving this system yields closed-form expressions for the update coefficients:
\begin{equation}
    a_t = \frac{\sqrt{1-\bar{\alpha}_{t-1}}}{\sqrt{1-\bar{\alpha}_t}},
    \label{eq:a}
\end{equation}
\begin{equation}
    b_t = \sqrt{\bar{\alpha}_{t-1}} - \frac{\sqrt{\bar{\alpha}_t (1-\bar{\alpha}_{t-1})}}{\sqrt{1-\bar{\alpha}_t}},
    \label{eq:b}
\end{equation}
\begin{equation}
    \gamma_t = \gamma_{t-1} = \frac{\sqrt{\bar{\alpha}_{N_s}}}{\sqrt{1 - \bar{\alpha}_{N_s}}}.
    \label{eq:c}
\end{equation}
During the sampling stage, the latent variable is iteratively updated according to the model-predicted conditional distribution $p_\theta(z_{t-1} \mid z_t, \tilde{z}_0, z_c)$. In particular, the deterministic update rule can be written as:
$
    z_{t-1} = a_t z_t + b_t \tilde{z}_0,
$
where $\tilde{z}_0 = \tilde{z}_0(z_t,z_c,t,o; \theta)$ denotes the predicted clean latent estimated from the current noisy latent $z_t$. The overall sampling procedure is summarized in Algorithm~\ref{alg:1}.


\subsection{Training Process}



During training, the diffusion sampling objective at step $t$ is defined as  
\begin{equation}
    L_{d,t}=\mathbb{E}_{z_0,z_c,\epsilon}\big\| (a_tz_t+b_tz_0) - (a_tz_t+b_t\hat z_0)\big\|^2,
\end{equation}
which can be simplified for $n<N_s$ as  
\begin{equation}
    L_{d} = \left(\frac{b_t\sqrt{1-\bar{\alpha}_t}}{\sqrt{\bar{\alpha}_t}-\sqrt{1-\bar{\alpha}_t}\gamma_t}\right)^2
    \mathbb{E}_{z_0,z_c,\epsilon}\big\|\epsilon-\epsilon_\theta(z_t,z_c, o,t)\big\|^2 .
\end{equation}
For training stability, we omit the coefficient term in $L_{d}$ and directly use the simplified diffusion loss form during training.

Throughout the entire training process, the backbone of the pretrained Stable Diffusion model remains frozen. Our training procedure exclusively optimizes the parameters of the JSCC module and the ControlNet. The process consists of two stages:

In the first stage, we jointly train the JSCC module and the ControlNet. The objective is twofold: 1) to train the ControlNet to effectively inject the decoded structural condition $\mathbf{z}_c$ into the frozen diffusion backbone, and 2) to ensure the JSCC module accurately reconstructs the latent code. The training loss therefore combines the diffusion objective with a latent-domain MSE term:
\begin{equation}
\begin{aligned}
    L_{1} = &\mathbb{E}_{z_0,z_c,\epsilon}\big\|\epsilon-\epsilon_\theta(z_t,z_c,o,t)\big\|^2 \\
    &+ \lambda_{D}\,\mathrm{MSE}(z,z_c).
\end{aligned}
\end{equation}
Here, $\lambda_D$ is a weighting factor that balances the primary generative task (the diffusion loss) with the latent reconstruction fidelity (the MSE loss).

In the second stage, we place a greater emphasis on the perceptual quality of the final reconstructed image and therefore introduce additional loss components.
Following (\ref{eq:pred_z0}), the predicted latent $\hat z_0$ is obtained and then decoded into the pixel domain to generate the reconstructed image $\tilde{\mathbf{s}}$. To further enhance perceptual quality, we introduce an additional optimization step based on a weighted combination of LPIPS and MSE losses. The overall training objective in the second stage is therefore defined as
\begin{equation}
\begin{aligned}
    L_{2} = &\mathbb{E}_{z_0,z_c,\epsilon}\big\|\epsilon-\epsilon_\theta(z_t,z_c,o,t)\big\|^2 + \lambda_{D}\,\mathrm{MSE}(z,z_c) \\
    &+ \lambda_{M}\,\mathrm{MSE}(\mathbf{s},\tilde{\mathbf{s}}) 
    + \lambda_{L}\,\mathrm{LPIPS}(\mathbf{s},\tilde{\mathbf{s}}),
\end{aligned}
\end{equation}
where the first two terms maintain the generative consistency established in the first stage, while the latter two terms jointly optimize pixel-level fidelity and perceptual realism in the reconstructed image. The hyperparameters $\lambda_{M}$ and $\lambda_{L}$ balance the fidelity-oriented and perceptual-oriented objectives. 

\section{Experiment Results}

In this section, we presents a series of experiments designed to evaluate the performance of the proposed strategy under various scenarios.

\subsection{Experimental Setup}

We adopt MambaJSCC~\cite{10901192} as the JSCC scheme for transmitting either the image or its latent representation over the wireless channel. Stable Diffusion v2.1 is employed as the backbone diffusion model for image generation and refinement. In addition, the MLLM GPT-4o is used as the visual-language component to generate textual descriptions for semantic conditioning.

For training, we use the LSDIR~\cite{li2023lsdir} and Flickr2W~\cite{liu2020unified} datasets. All images are resized and center cropped to $256\times256$ during preprocessing. To enable classifier-free guidance, we randomly drop 10\% of the text prompt during training.

For the two training stage, the training configurations are as follows
In the first stage, we set $\lambda_M = \lambda_L = 0$ and $\lambda_D = 1$, allowing the model to focus solely on latent-space reconstruction. In the second stage, all three loss weights are activated, with $\lambda_M = 10$ and $\lambda_L = \lambda_D = 1$, emphasizing both perceptual fidelity and reconstruction accuracy in the final image domain.

\begin{figure*}[htbp]
    \centering
    \subfigure[LPIPS, AWGN]{\includegraphics[width=0.245\linewidth]{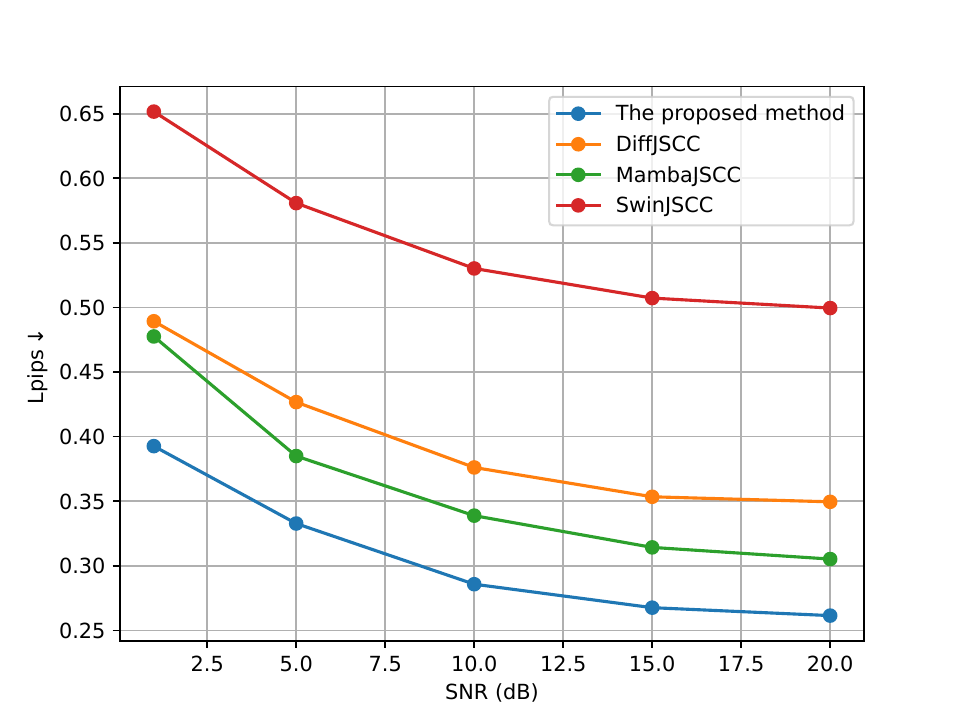}}
    \subfigure[DISTS, AWGN]{\includegraphics[width=0.245\linewidth]{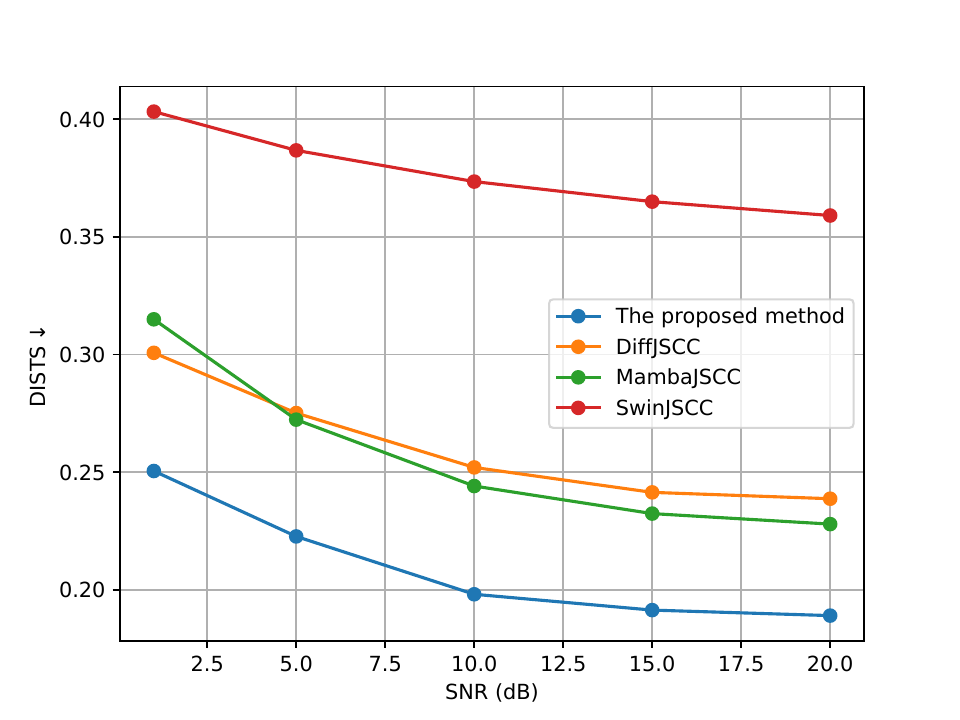}}
    \subfigure[FID, AWGN]{\includegraphics[width=0.245\linewidth]{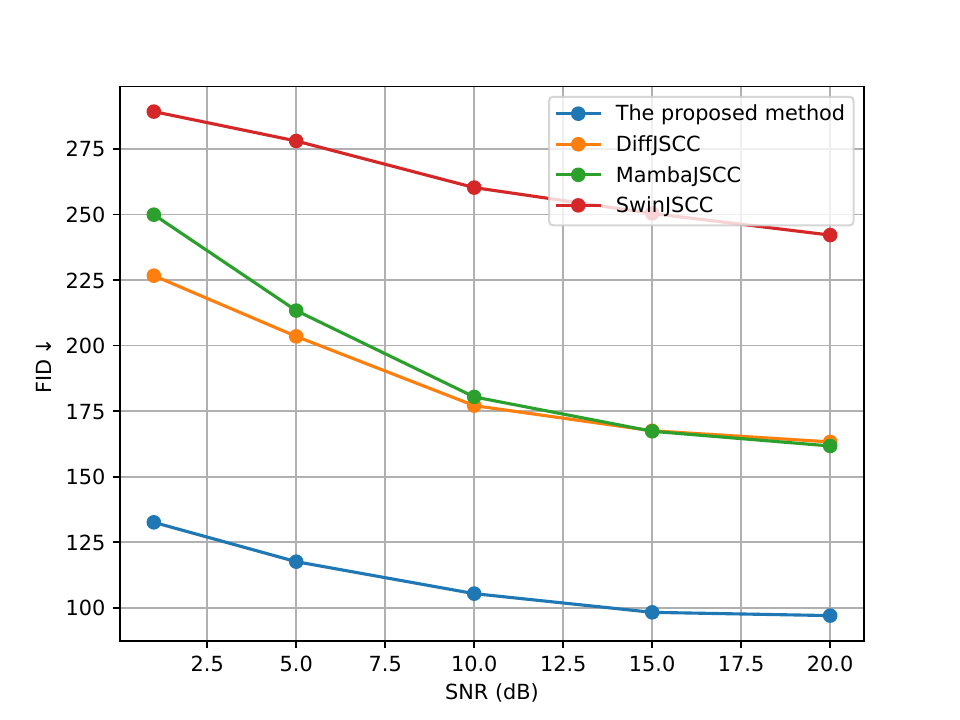}}
    \subfigure[KID, AWGN]{\includegraphics[width=0.245\linewidth]{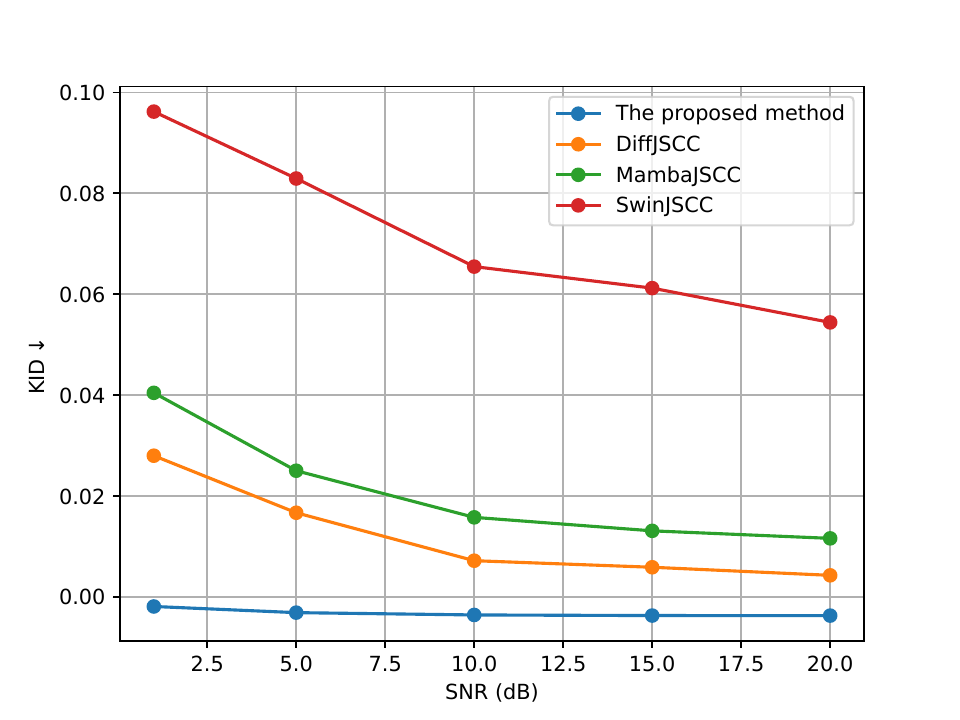}}\\
    \subfigure[LPIPS, Rayleigh]{\includegraphics[width=0.245\linewidth]{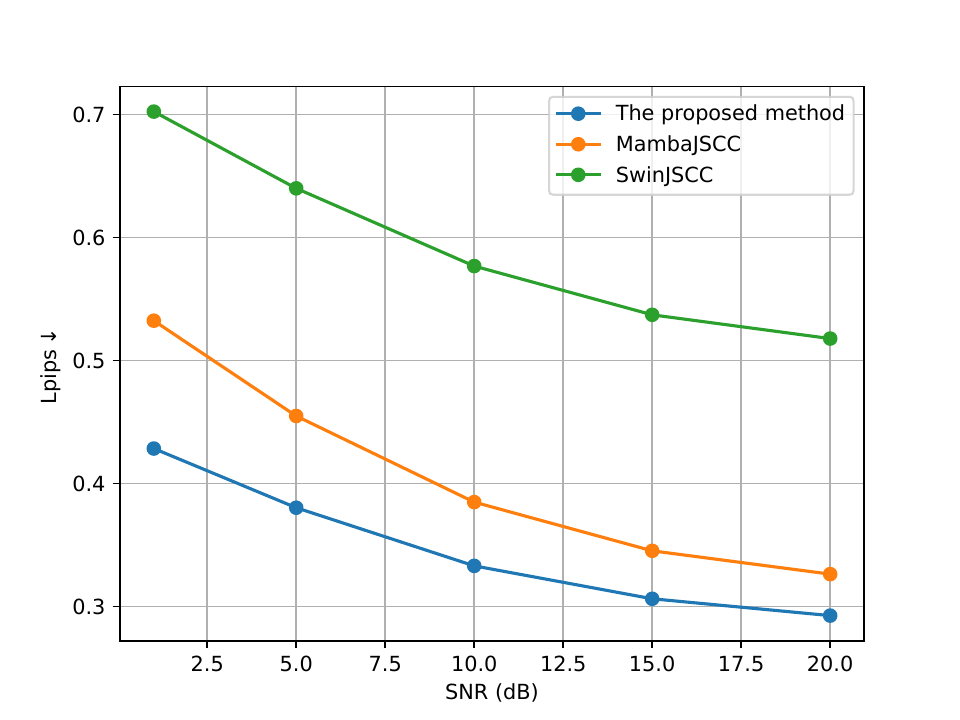}}
    \subfigure[DISTS, Rayleigh]{\includegraphics[width=0.245\linewidth]{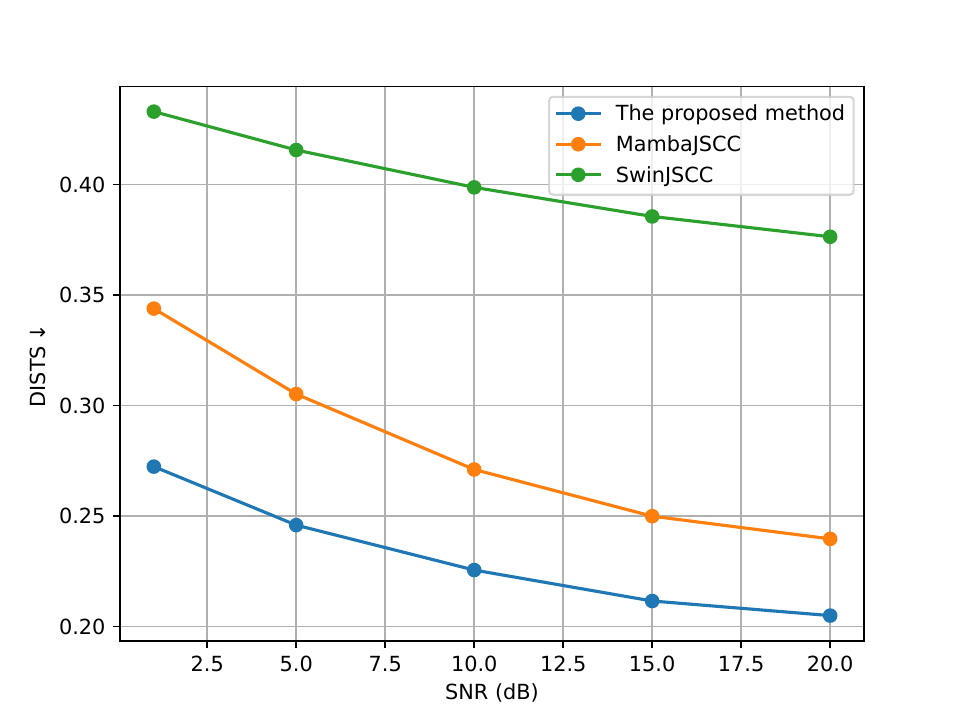}}
    \subfigure[FID, Rayleigh]{\includegraphics[width=0.245\linewidth]{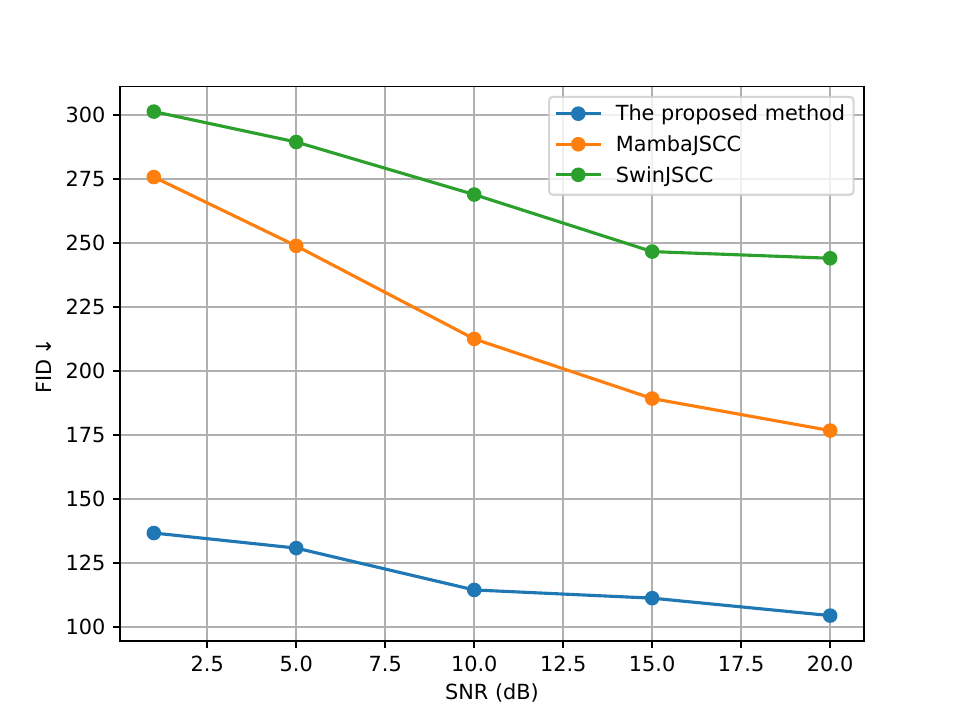}}
    \subfigure[KID, Rayleigh]{\includegraphics[width=0.245\linewidth]{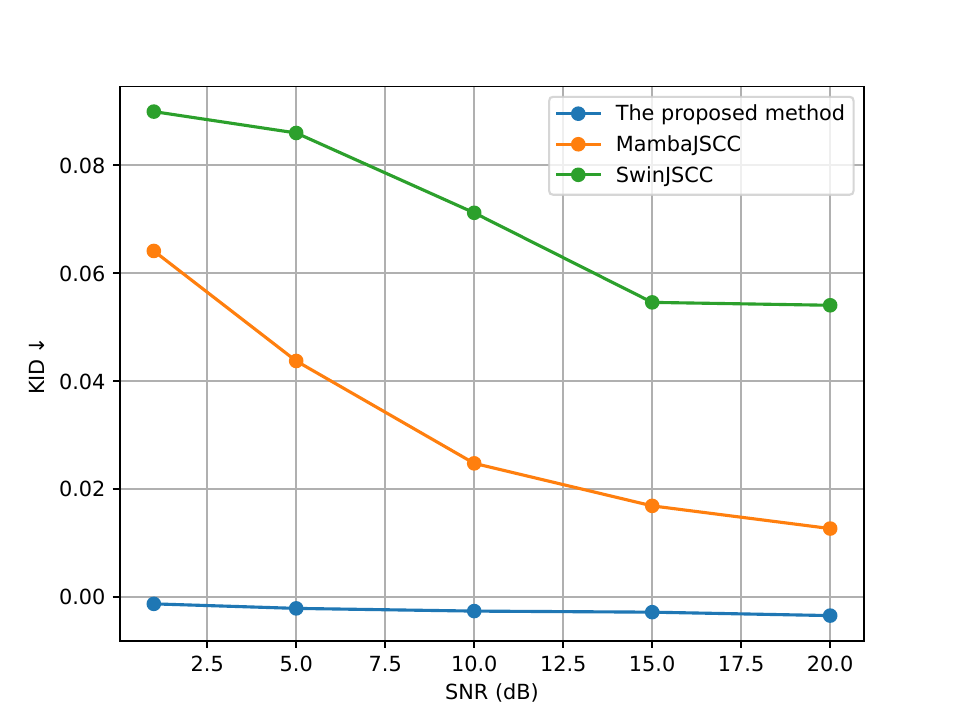}}
    \caption{Performance of different models on DIV2k dataset under AWGN and Rayleigh channel across different SNR. The CBR is set to 0.00326.}
    \label{fig:div2k_snr}
\end{figure*}

\begin{figure*}[htbp]
    \centering
    \subfigure[LPIPS, AWGN]{\includegraphics[width=0.245\linewidth]{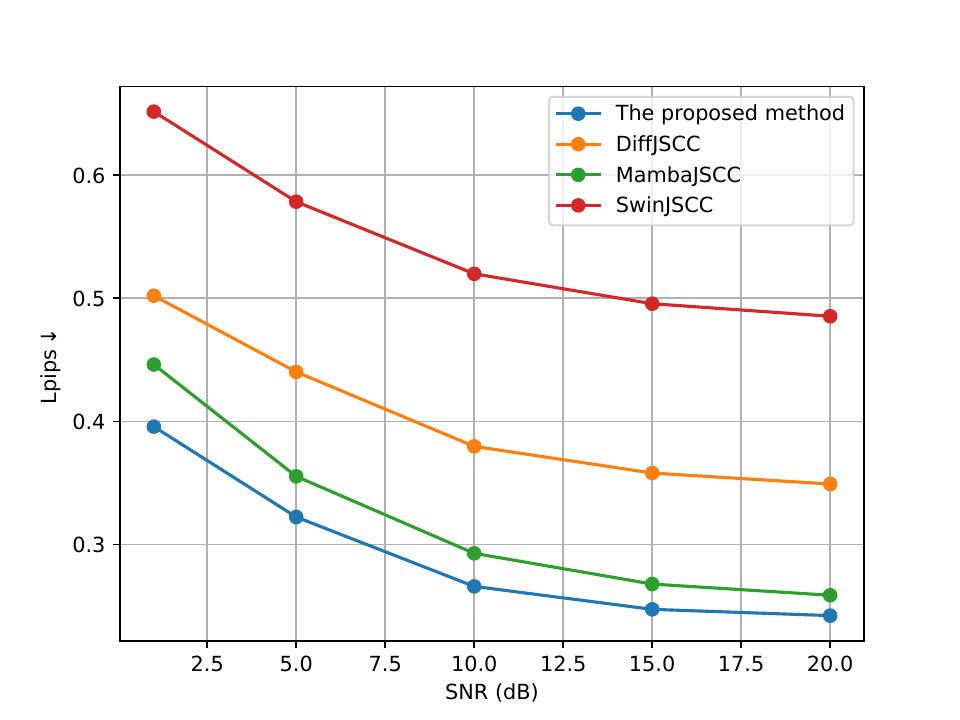}}
    \subfigure[DISTS, AWGN]{\includegraphics[width=0.245\linewidth]{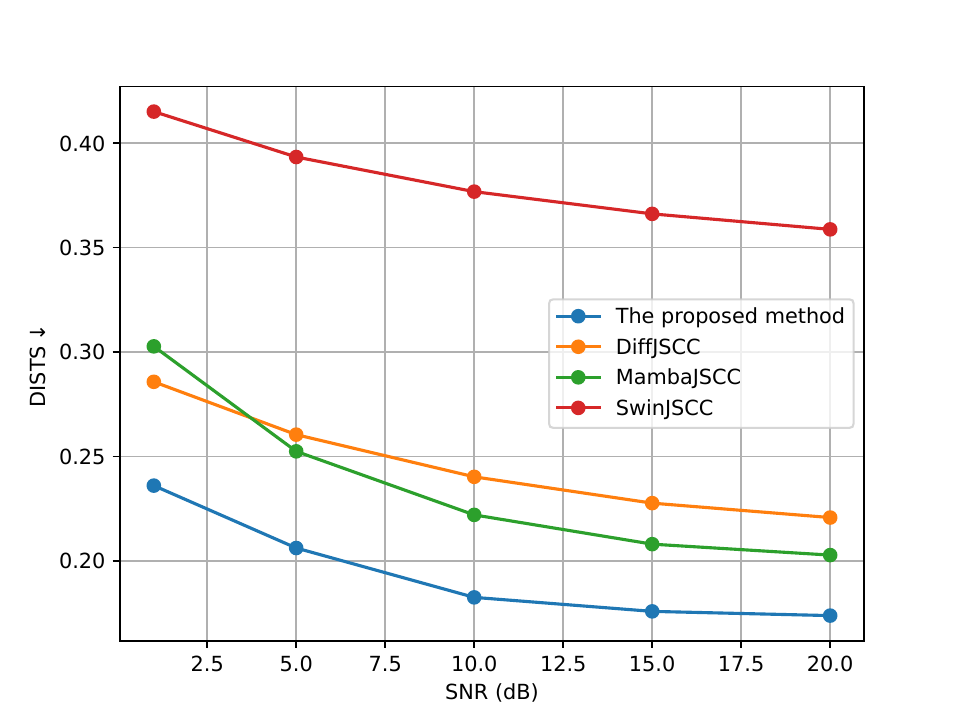}}
    \subfigure[FID, AWGN]{\includegraphics[width=0.245\linewidth]{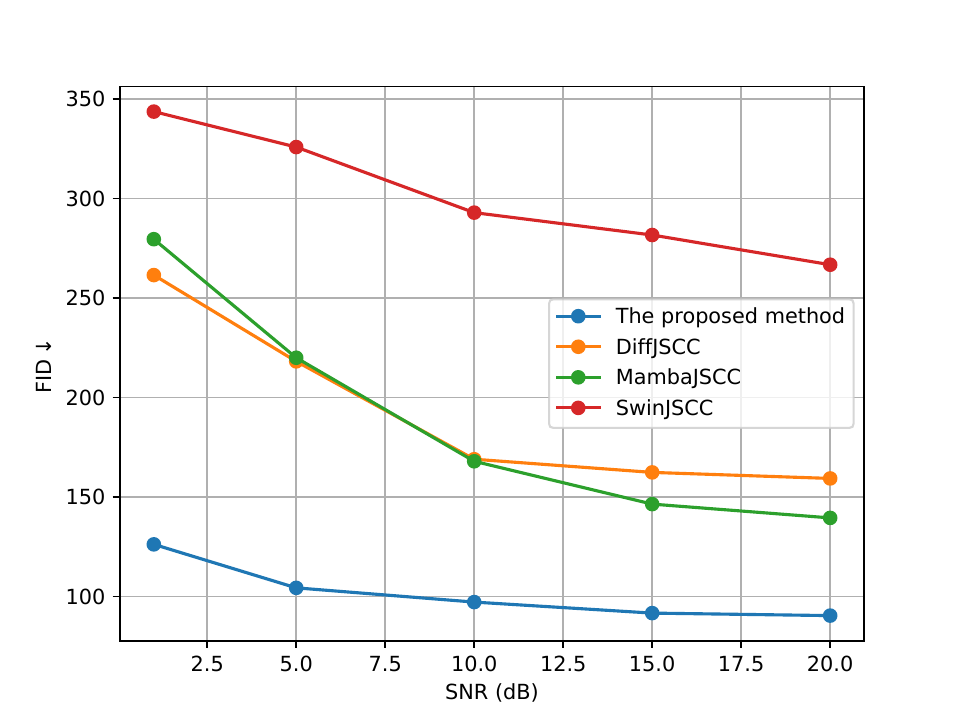}}
    \subfigure[KID, AWGN]{\includegraphics[width=0.245\linewidth]{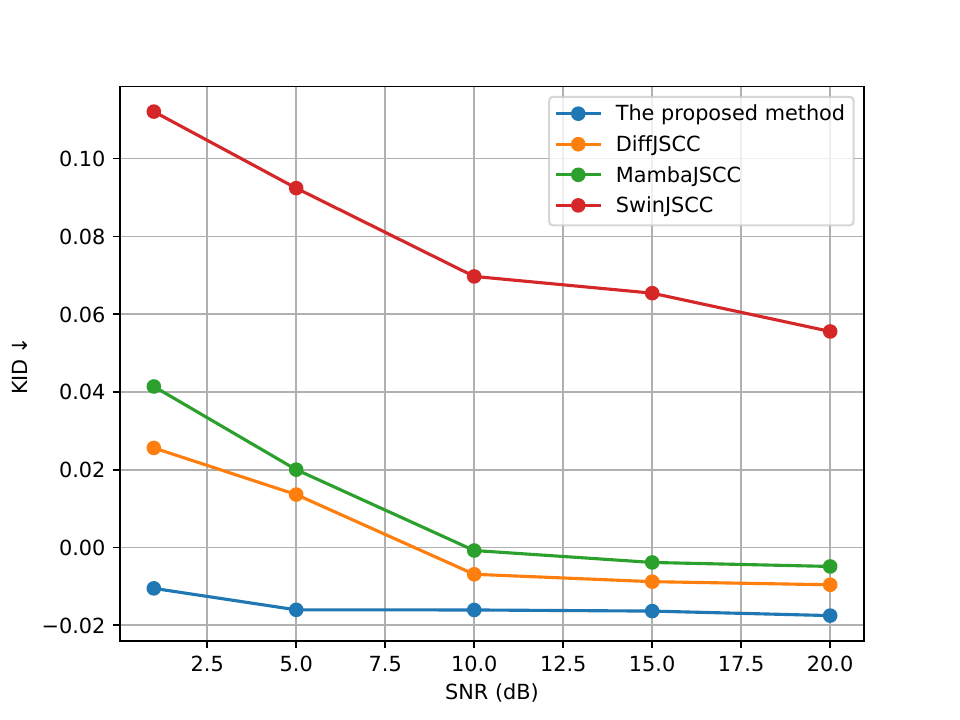}}
    \subfigure[LPIPS, Rayleigh]{\includegraphics[width=0.245\linewidth]{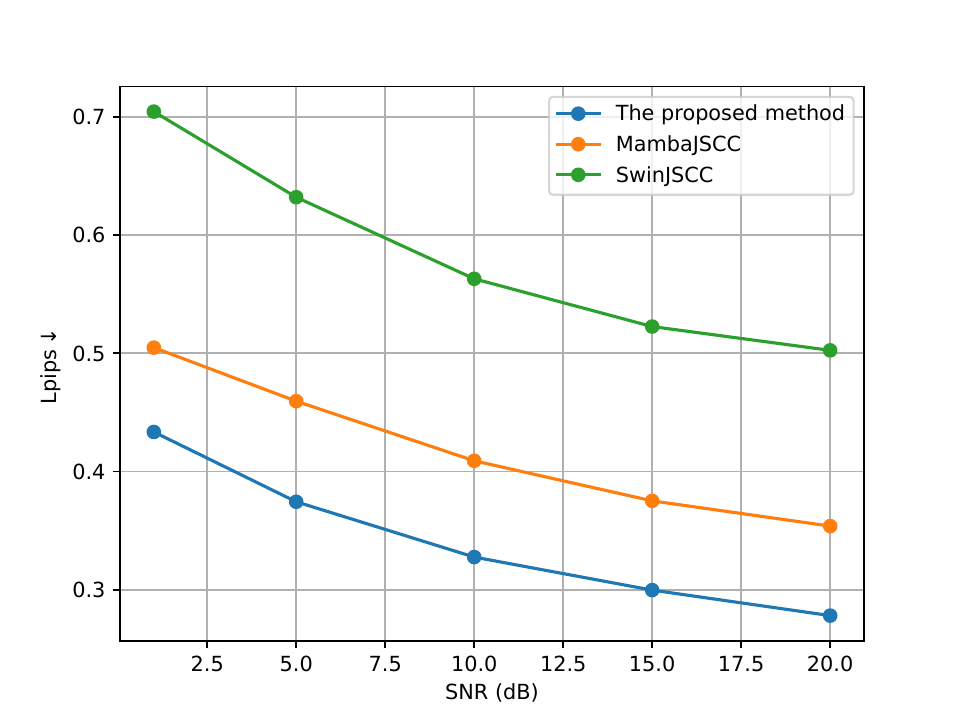}}
    \subfigure[DISTS, Rayleigh]{\includegraphics[width=0.245\linewidth]{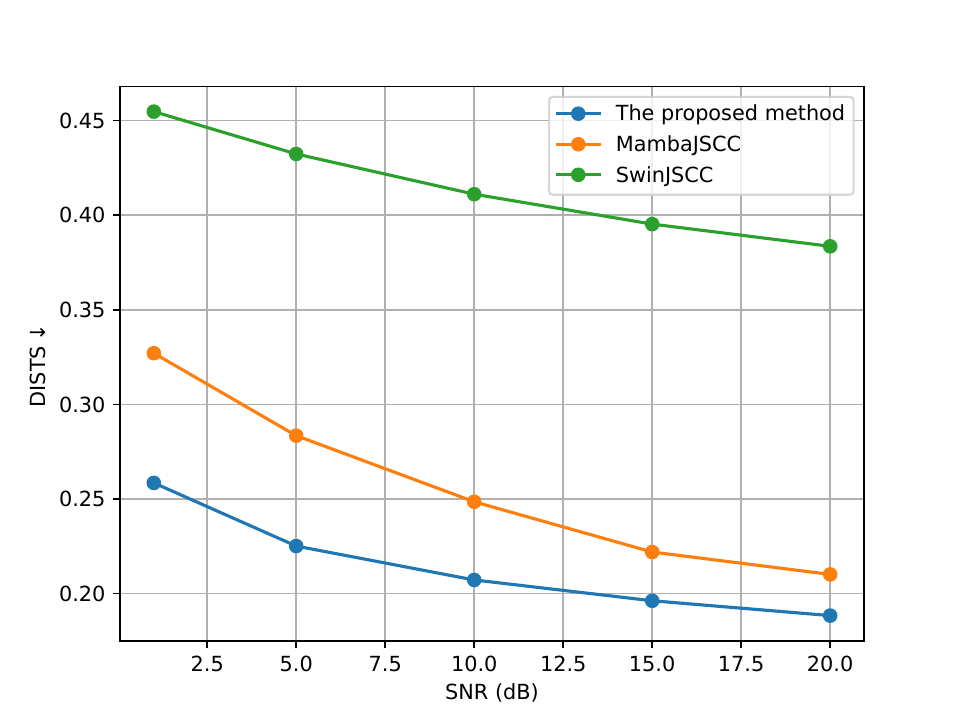}}
    \subfigure[FID, Rayleigh]{\includegraphics[width=0.245\linewidth]{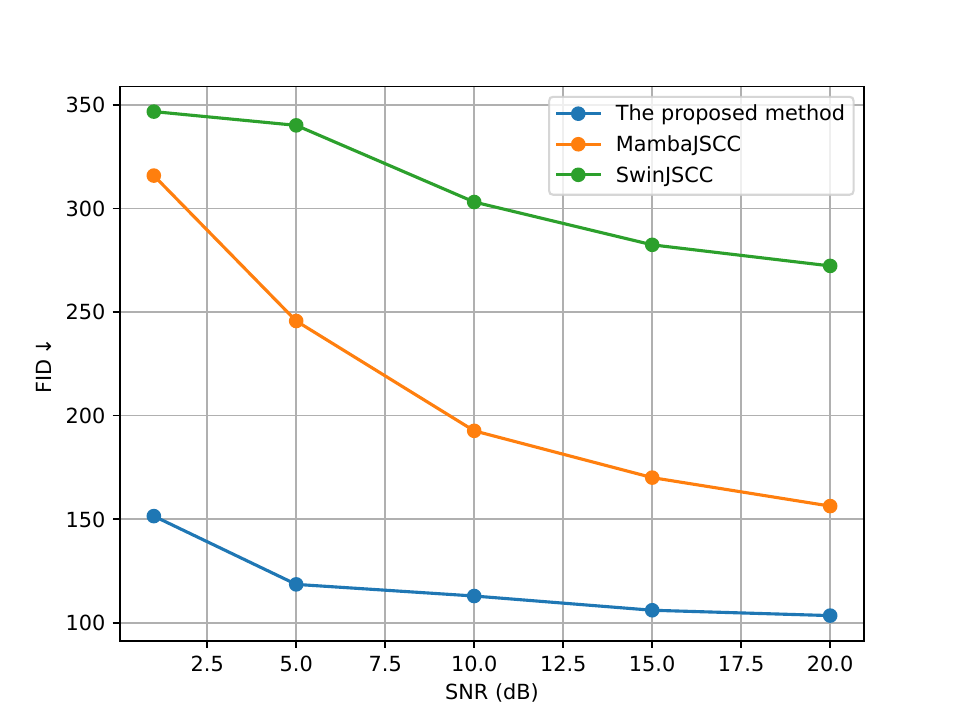}}
    \subfigure[KID, Rayleigh]{\includegraphics[width=0.245\linewidth]{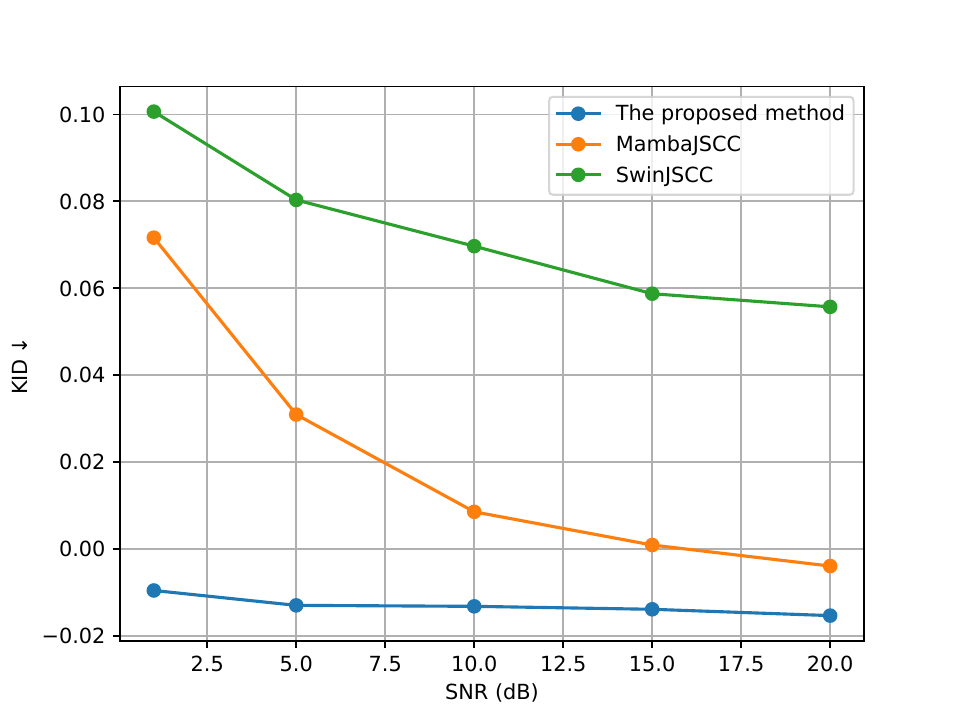}}
    \caption{Performance of different models on Kodak dataset under AWGN and Rayleigh channel across different SNR. The CBR is set to 0.00326.}
    \label{fig:kodak_snr}
\end{figure*}

For CBRs of $0.0020$, $0.0033$, $0.0059$, and $0.011$, the diffusion sampling starts from steps $N_s$ is set to $600, 500, 400$ and $300$, respectively.

For evaluation, we adopt the DIV2K~\cite{agustsson2017ntire} and Kodak~\cite{kodakdataset} datasets. The DIV2K dataset consists of 900 high-quality images with $2\text{K}$ resolution, covering diverse real-world scenes. Only the test split is used in our experiments.The Kodak dataset contains 24 widely adopted benchmark images, commonly used for evaluating image compression and reconstruction methods. 
We compare the proposed approach with state-of-the-art JSCC methods, including MambaJSCC~\cite{10901192}, SwinJSCC~\cite{yang2024swinjscc}, and the diffusion-based model DiffJSCC~\cite{yang2024diffusion}. For fair comparison, MambaJSCC is trained using a hybrid loss combining MSE and LPIPS, while SwinJSCC is trained solely with MSE, following the settings in their respective original works. All models are optimized using the AdamW optimizer with a learning rate of $10^{-4}$ until convergence, and training is performed on NVIDIA A40 GPUs.
The reconstruction quality is evaluated using both perceptual and distributional metrics, including the LPIPS~\cite{zhang2018unreasonable}, DISTS~\cite{ding2020image}, Fréchet Inception Distance (FID)~\cite{heusel2017gans}, and Kernel Inception Distance (KID)~\cite{binkowski2018demystifying}.

\subsection{Experimental Results}
\begin{figure*}[htbp]
    \centering
    \subfigure[LPIPS, AWGN]{\includegraphics[width=0.245\linewidth]{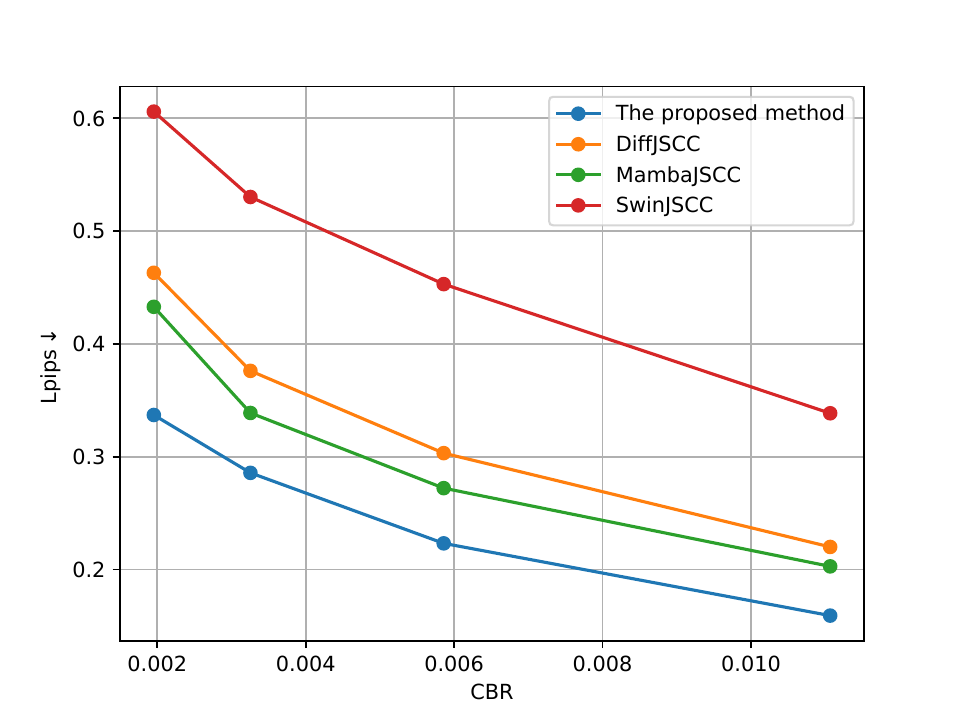}}
    \subfigure[DISTS, AWGN]{\includegraphics[width=0.245\linewidth]{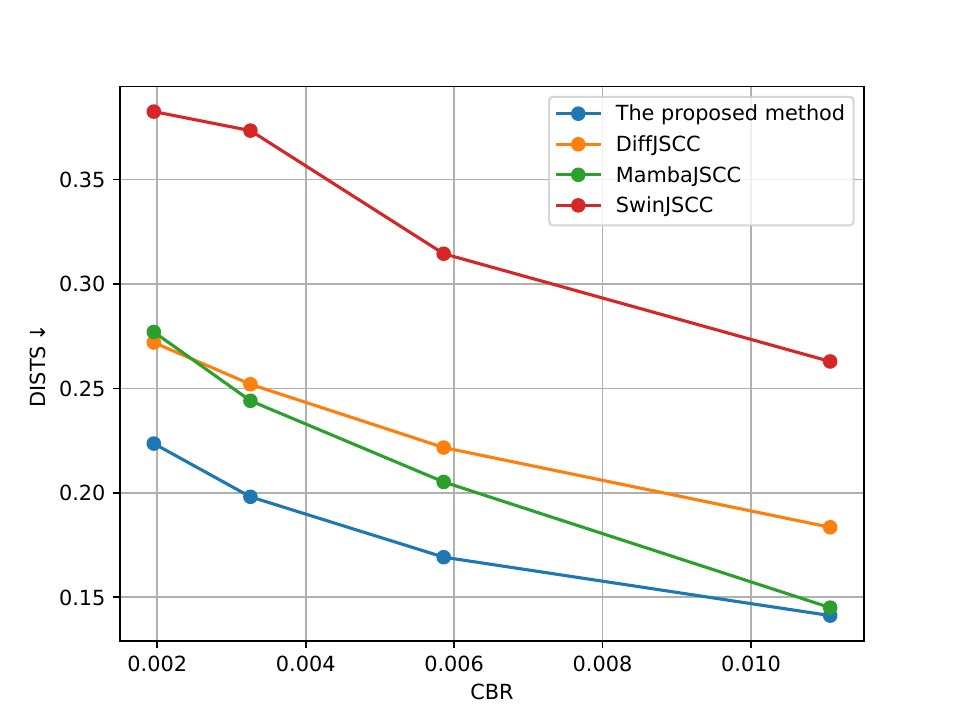}}
    \subfigure[FID, AWGN]{\includegraphics[width=0.245\linewidth]{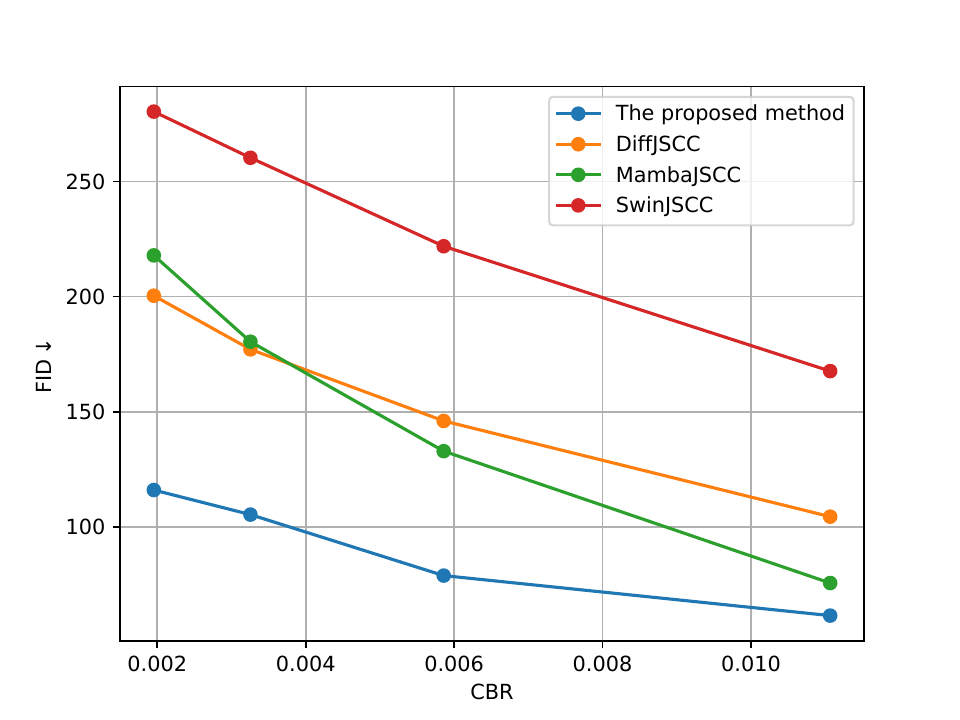}}
    \subfigure[KID, AWGN]{\includegraphics[width=0.245\linewidth]{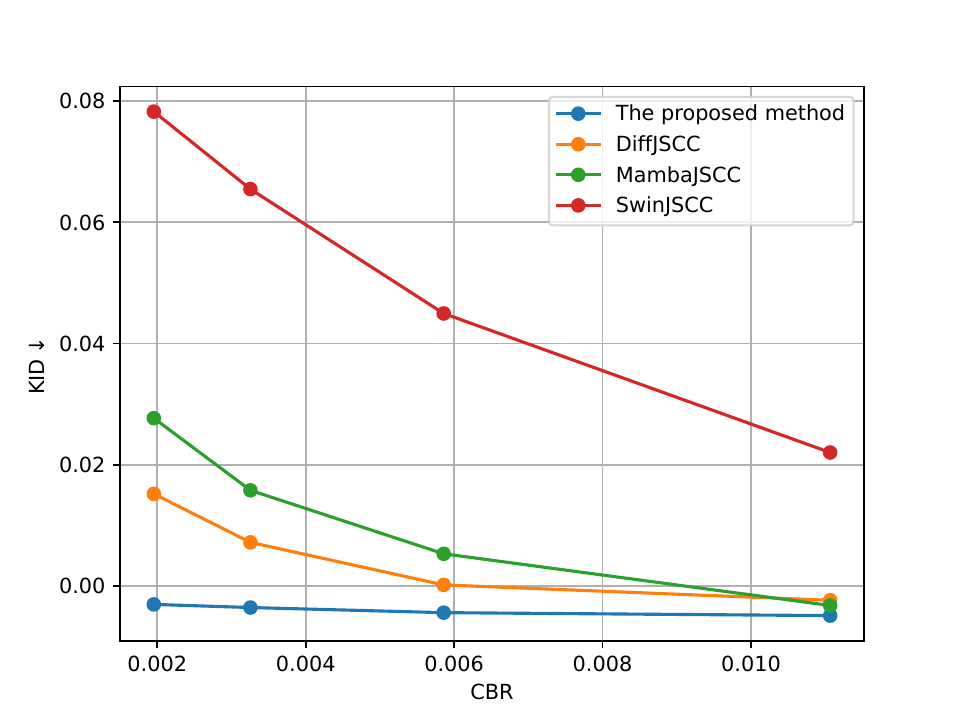}}
    \subfigure[LPIPS, Rayleigh]{\includegraphics[width=0.245\linewidth]{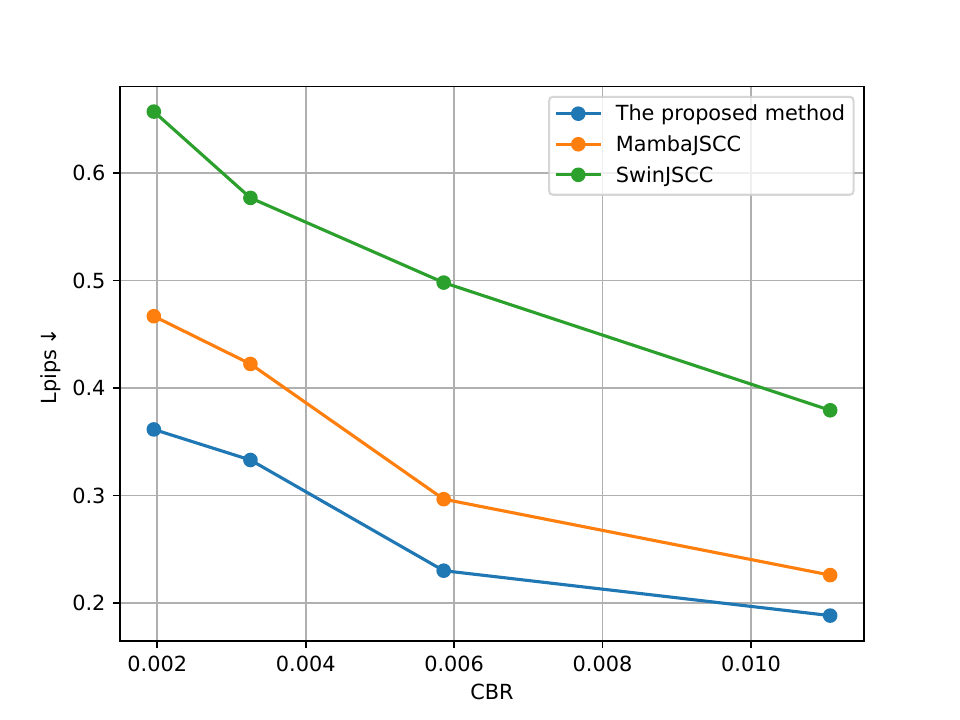}}
    \subfigure[DISTS, Rayleigh]{\includegraphics[width=0.245\linewidth]{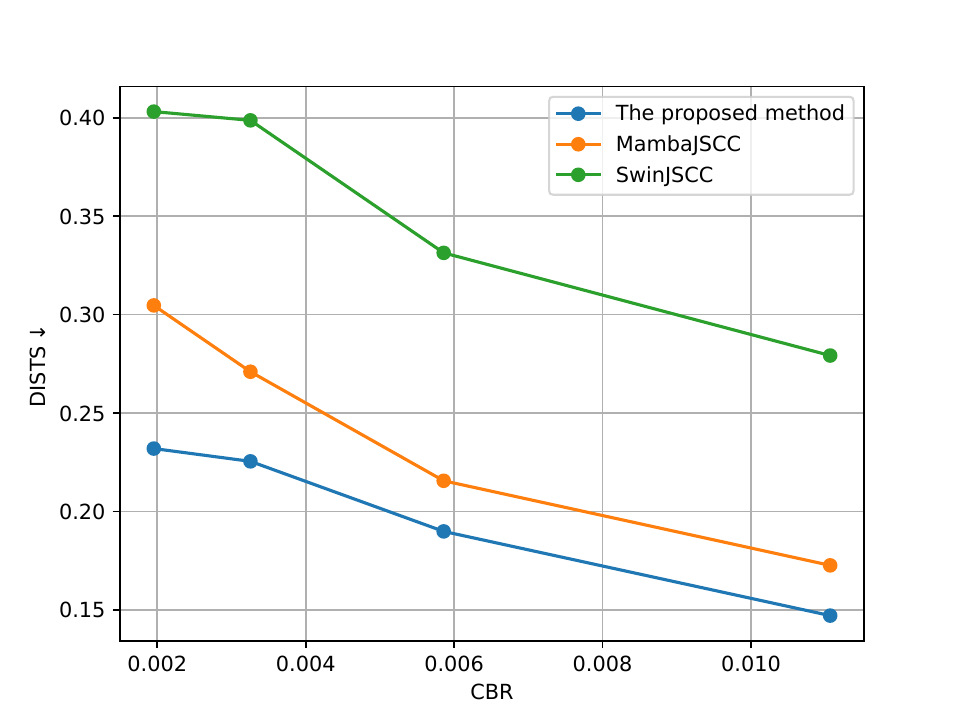}}
    \subfigure[FID, Rayleigh]{\includegraphics[width=0.245\linewidth]{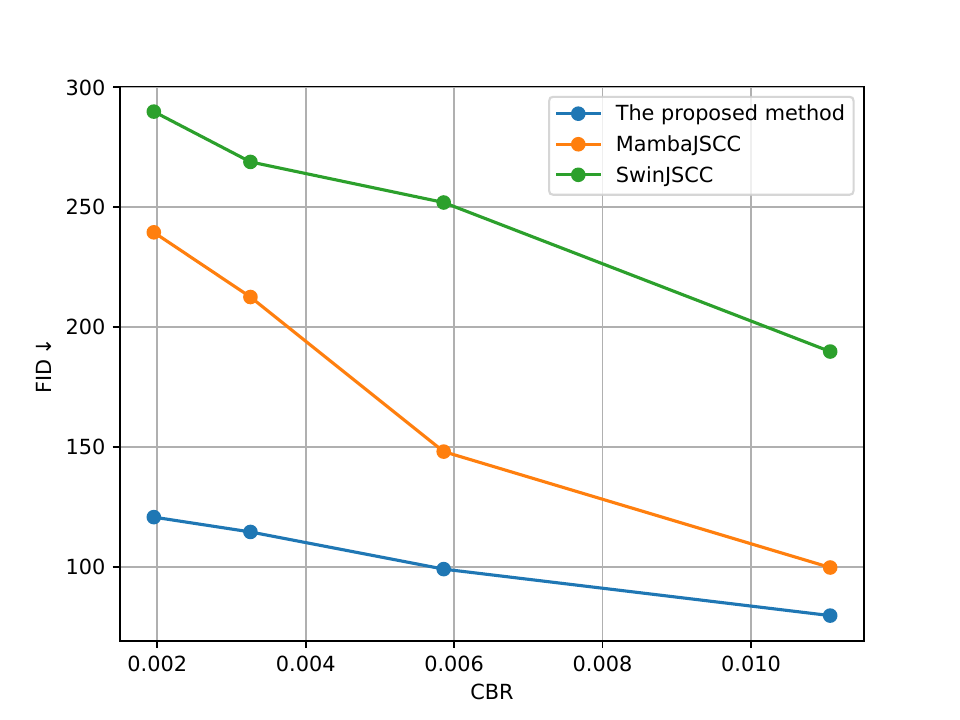}}
    \subfigure[KID, Rayleigh]{\includegraphics[width=0.245\linewidth]{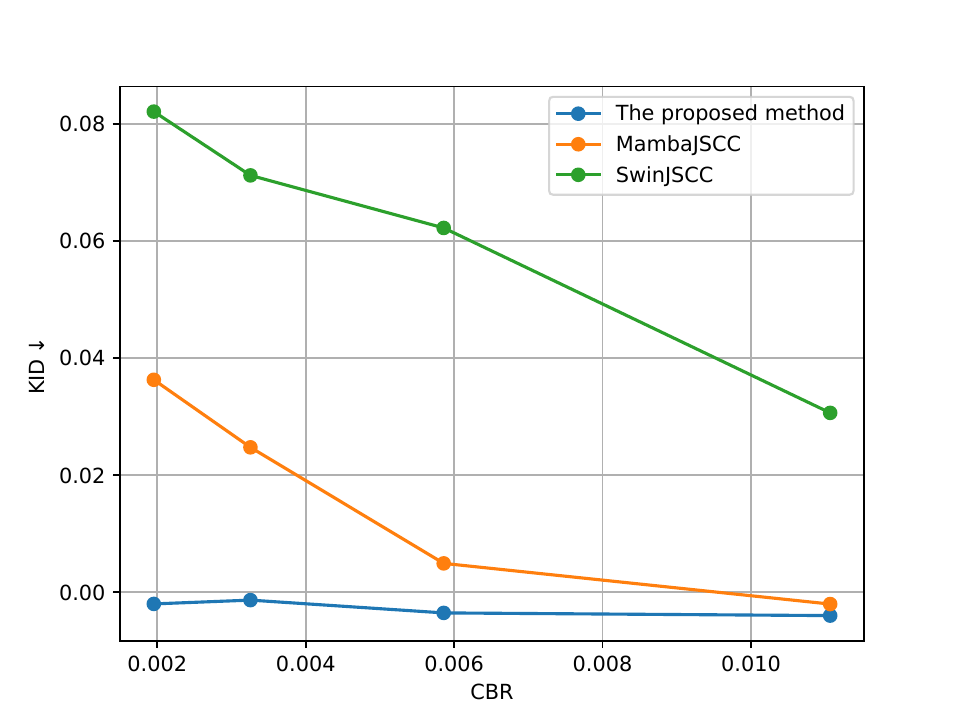}}
    \caption{Performance of different models on DIV2k dataset under AWGN and Rayleigh channel across different CBRs. The SNR is set to 10 dB.}
    \label{fig:div2k_cbr}
\end{figure*}

\begin{figure*}[htbp]
    \centering
    \subfigure[LPIPS, AWGN]{\includegraphics[width=0.245\linewidth]{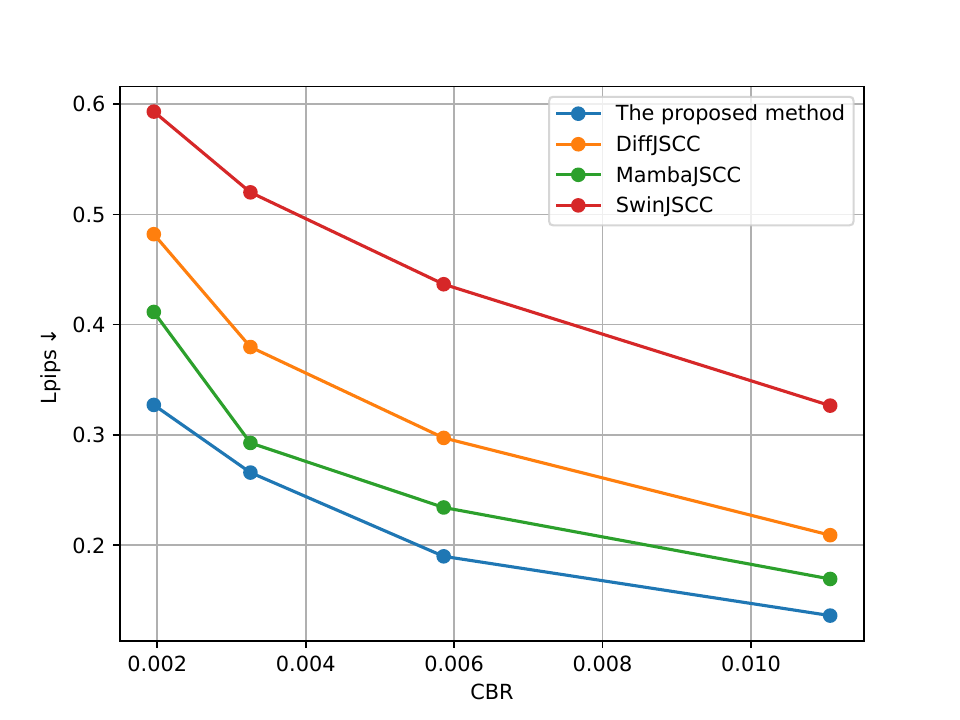}}
    \subfigure[DISTS, AWGN]{\includegraphics[width=0.245\linewidth]{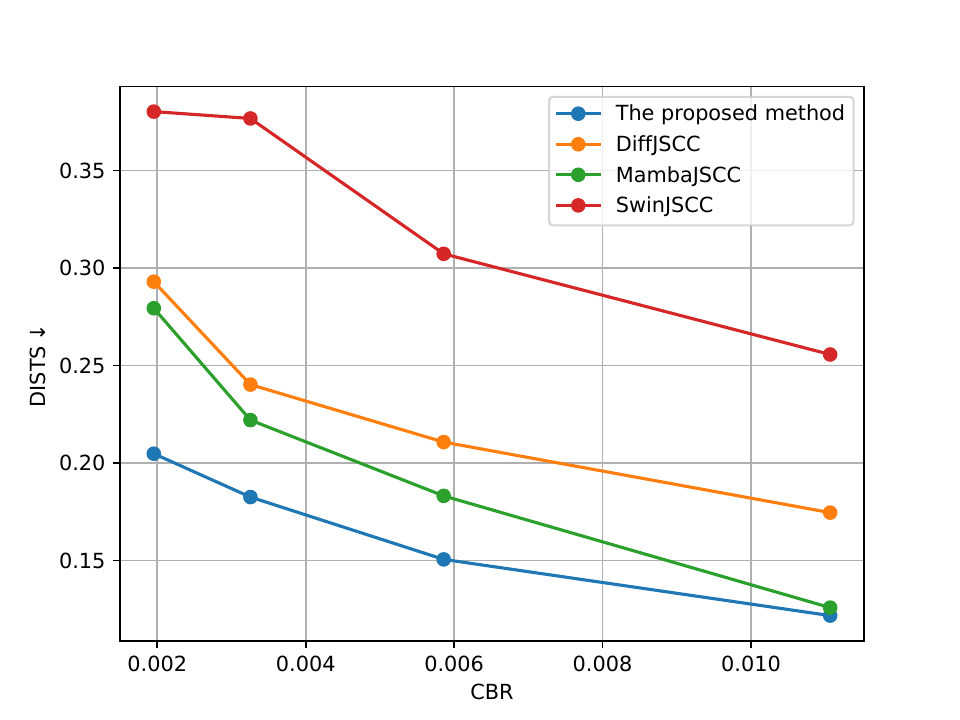}}
    \subfigure[FID, AWGN]{\includegraphics[width=0.245\linewidth]{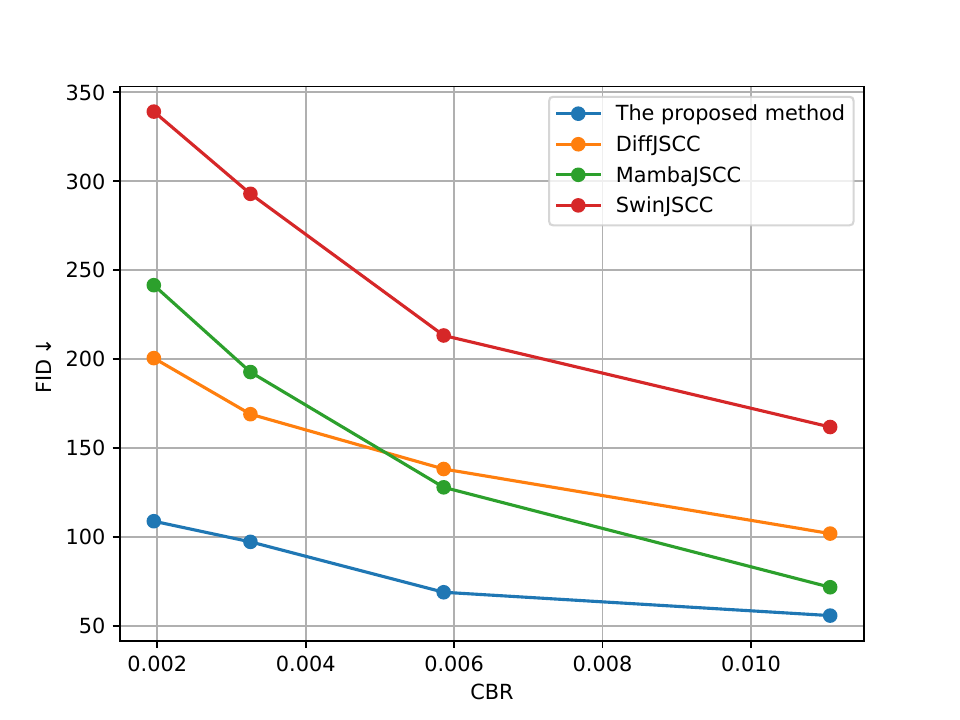}}
    \subfigure[KID, AWGN]{\includegraphics[width=0.245\linewidth]{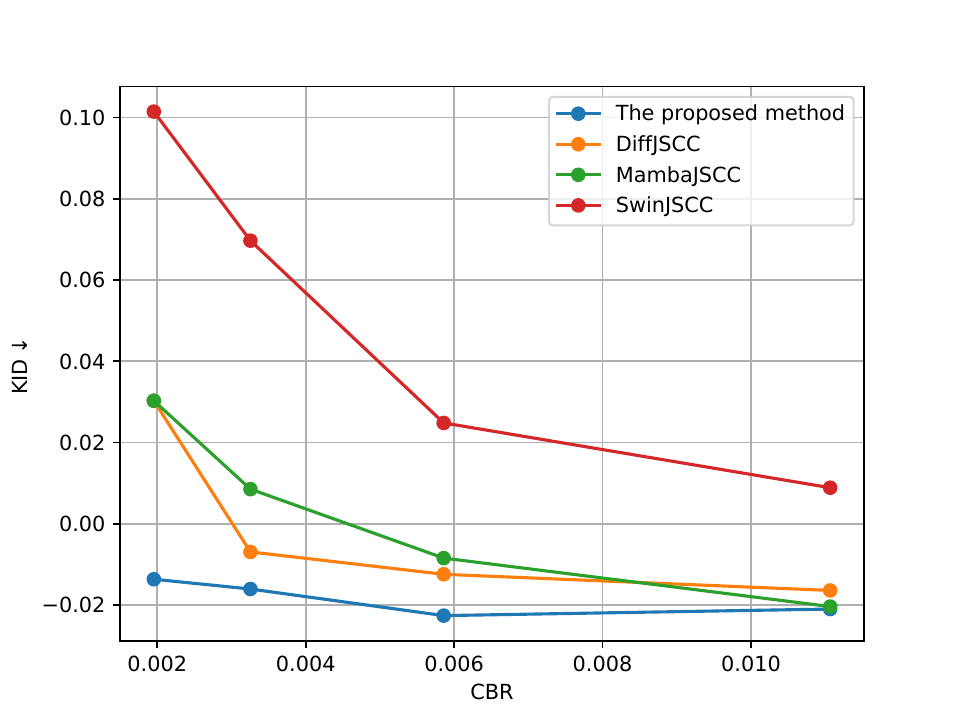}}
    \subfigure[LPIPS, Rayleigh]{\includegraphics[width=0.245\linewidth]{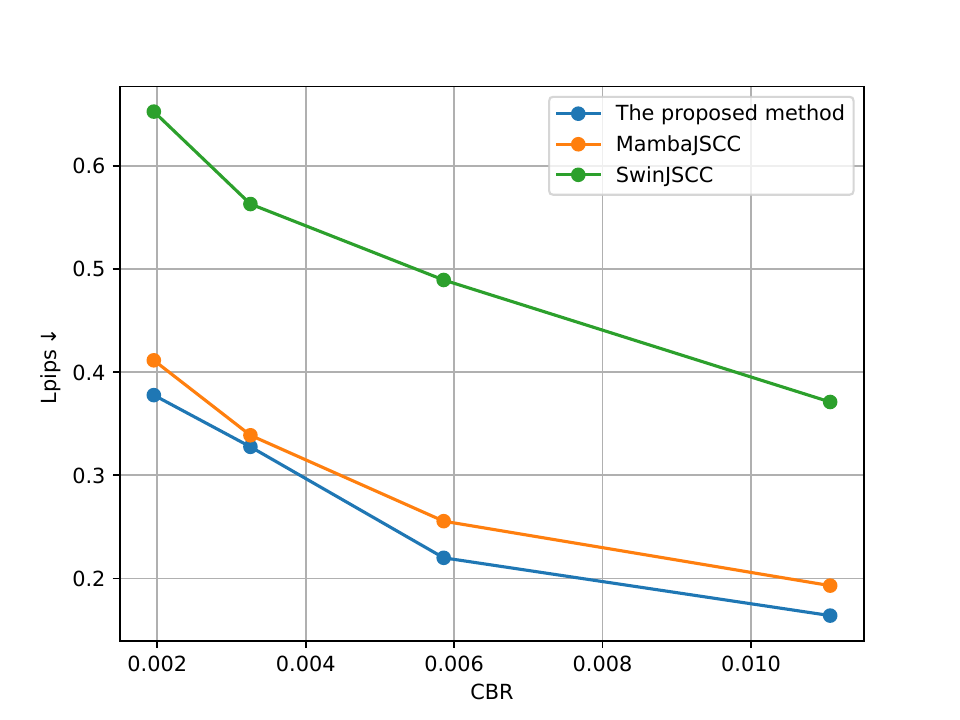}}
    \subfigure[DISTS, Rayleigh]{\includegraphics[width=0.245\linewidth]{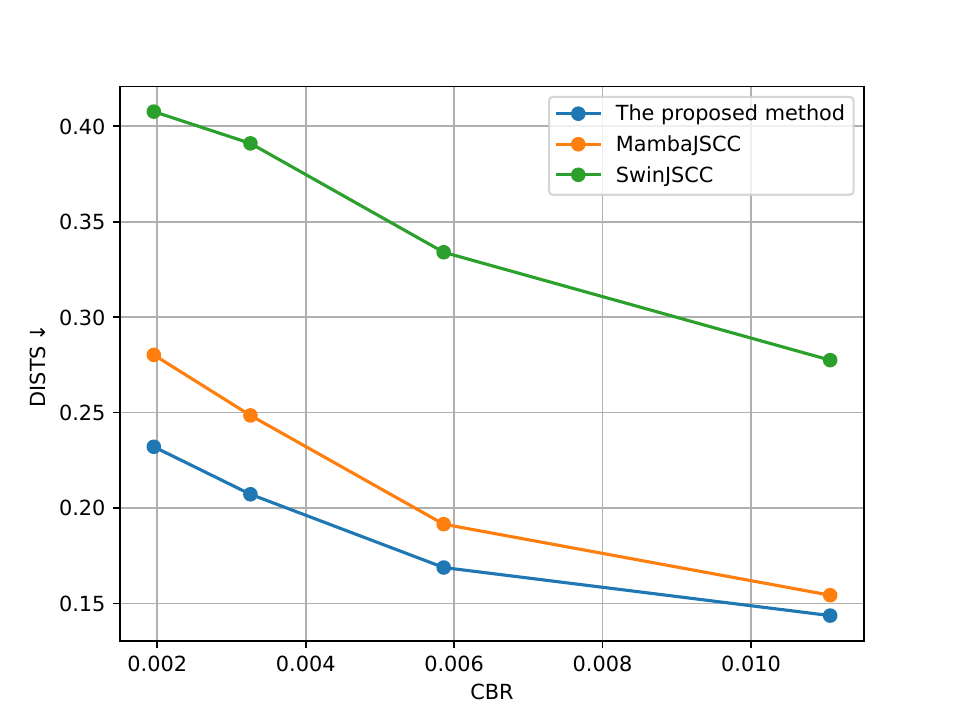}}
    \subfigure[FID, Rayleigh]{\includegraphics[width=0.245\linewidth]{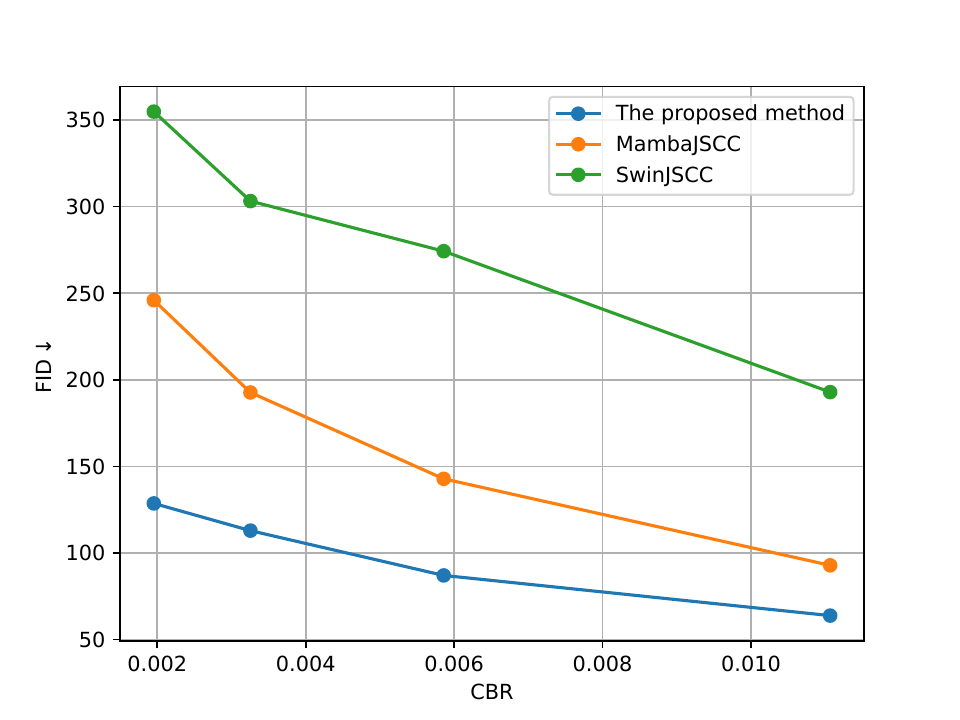}}
    \subfigure[KID, Rayleigh]{\includegraphics[width=0.245\linewidth]{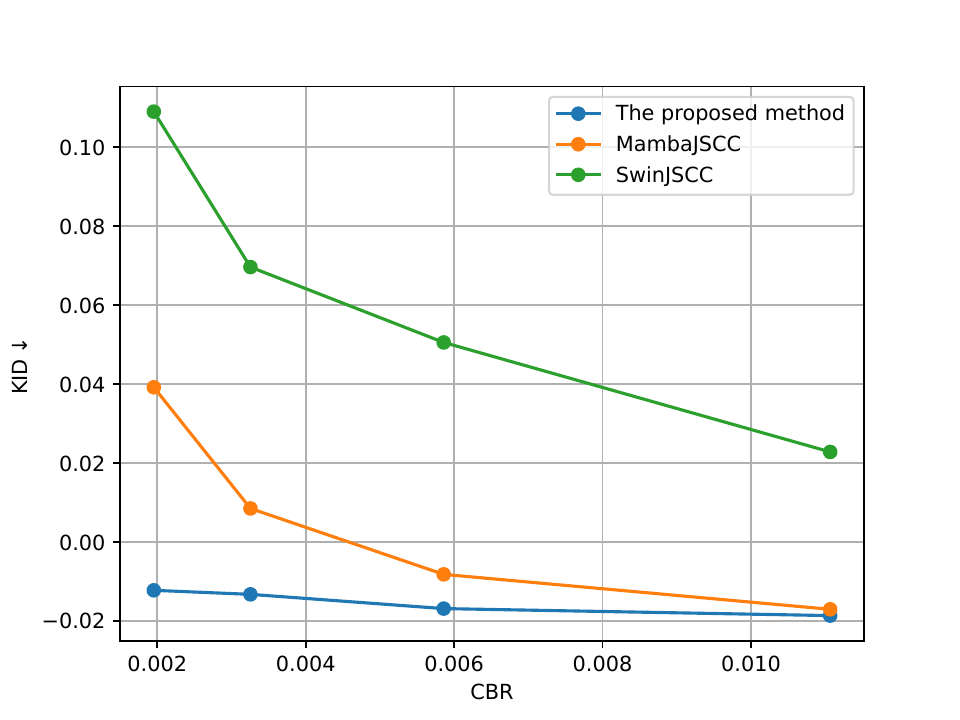}}
    \caption{Performance of different models on Kodak dataset under AWGN and Rayleigh channel across different CBRs. The SNR is set to 10 dB.}
    \label{fig:kodak_cbr}
\end{figure*}

Fig.~\ref{fig:div2k_snr} shows the LPIPS, DISTS, FID, and KID performance of various models on the DIV2K dataset under both AWGN and Rayleigh channels across different SNR levels. Since DiffJSCC only supports AWGN and slow-fading channels in its implementation, it is evaluated exclusively under AWGN and excluded from Rayleigh (fast-fading) comparisons for fairness. The proposed method consistently outperforms DiffJSCC and other JSCC baselines across all SNR levels and datasets. For example, at an SNR of $1$~dB under AWGN, the proposed model achieves a $0.085$ lower LPIPS than MambaJSCC and a $0.052$ lower DISTS than DiffJSCC, while under Rayleigh, it reduces FID by $133$ compared to MambaJSCC. SwinJSCC performs considerably worse than MambaJSCC, as it is trained solely with MSE loss.

Fig.~\ref{fig:kodak_snr} presents the corresponding results on the Kodak dataset. At an SNR of $10$~dB under Rayleigh, the proposed model attains $0.235$ lower LPIPS and $0.173$ lower DISTS than SwinJSCC. These trends are consistent with those observed on DIV2K: the proposed model delivers nearly identical performance across datasets and consistently surpasses existing JSCC baselines at all SNR levels. For instance, at $5$~dB under Rayleigh, it reduces LPIPS by $0.085$ and FID by $122$ compared to MambaJSCC.

\begin{figure*}[htbp]
    \centering
    \includegraphics[width=\linewidth]{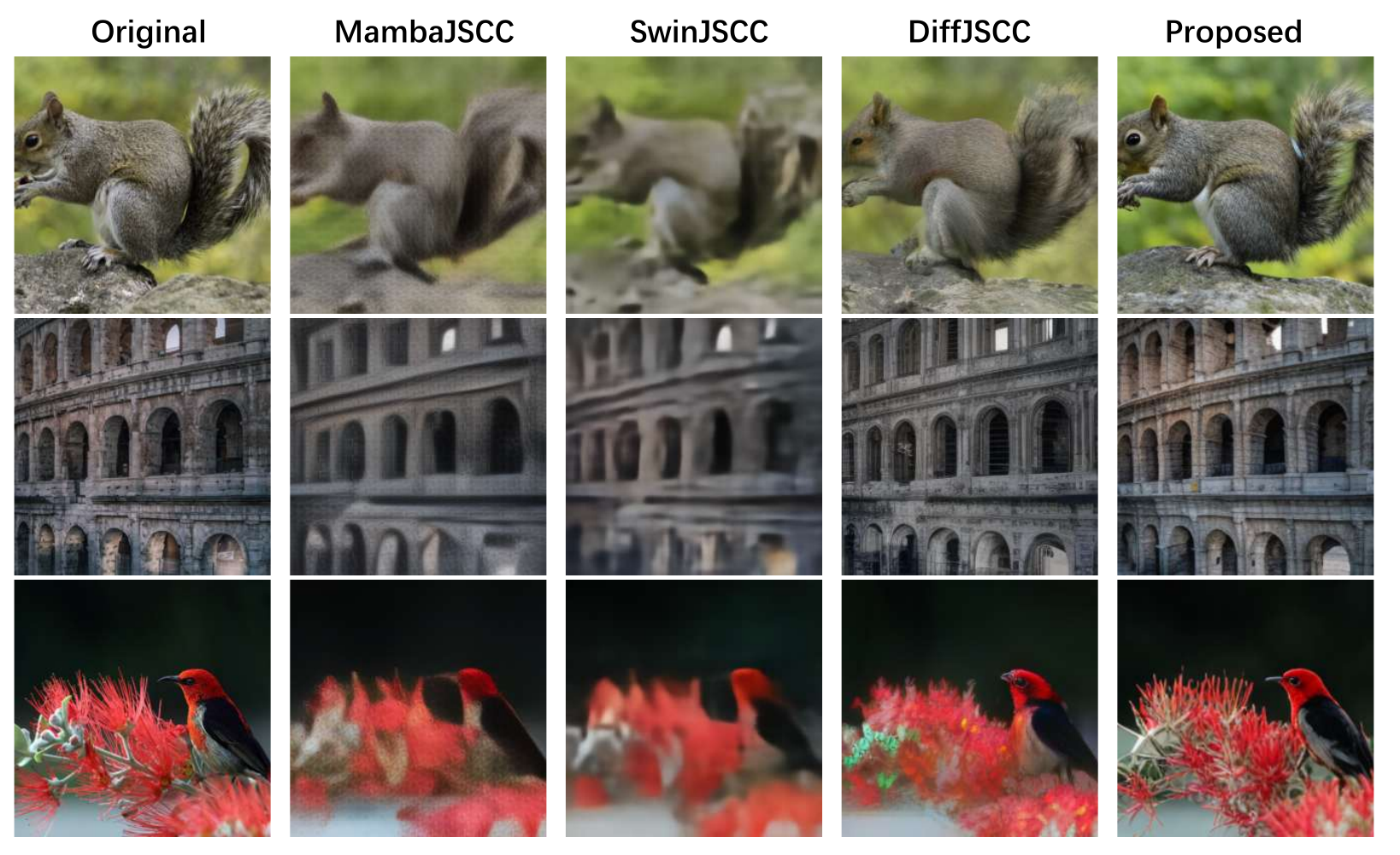}
    \caption{Examples of visual comparison between images reconstructed by different schemes under the AWGN channel at SNR=10 dB. The CBR is set to 0.0033.}
    \label{fig:visual}
\end{figure*}

We further evaluate the performance of the proposed models under varying CBRs to assess their robustness to different communication constraints. As shown in Fig.~\ref{fig:div2k_cbr}, both variants consistently outperform JSCC baselines in perceptual and distributional quality across all tested CBRs on the DIV2K dataset. For instance, at a CBR of $0.002$, the proposed method reduces LPIPS by $0.084$, DISTS by $0.054$, FID by $116$, and KID by $0.038$ compared to MambaJSCC. Even at a higher CBR of $0.011$, notable improvements remain, with LPIPS reduced by $0.040$, DISTS by $0.016$, and FID by $23$, demonstrating that the proposed approach effectively preserves perceptual and distributional quality under stringent channel constraints.

Fig.~\ref{fig:kodak_cbr} shows the results on the Kodak dataset. The trends mirror those observed on DIV2K: both proposed variants achieve nearly identical performance and consistently surpass MambaJSCC across all CBRs. At a CBR of $0.002$ under AWGN, the proposed model reduces LPIPS by $0.075$, DISTS by $0.074$, FID by $132$, and KID by $0.044$ compared to MambaJSCC. As the CBR increases, the performance gap narrows slightly, as higher channel capacity allows both models to retain more structural information. Overall, these results indicate that the proposed approach better preserves perceptual quality and semantic consistency, particularly under severe channel constraints. Notably, SwinJSCC generally underperforms relative to MambaJSCC across datasets and channels, and is therefore omitted from detailed comparisons for clarity.

Fig.~\ref{fig:visual} presents visual reconstruction results of different models over the AWGN channel at an SNR of 5 dB and a CBR of 0.0033. The proposed models produce clearer textures and more accurate structural details compared to DiffJSCC, while other JSCC baselines exhibit noticeable blurring and distortion.

\begin{figure}
    \centering
    \includegraphics[width=\linewidth]{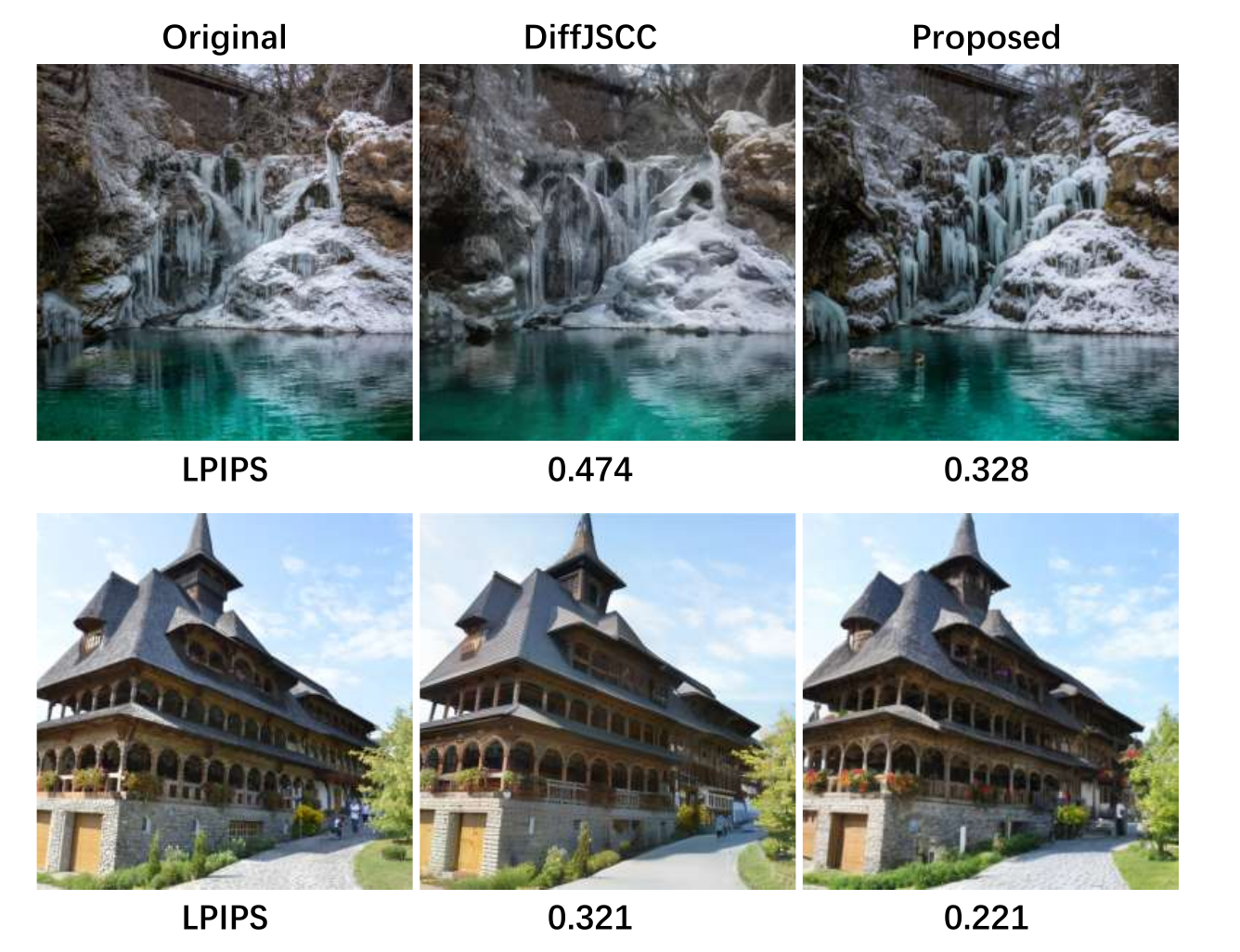}
    \caption{Comparison of reconstruction results under $512 \times 512$ resolutions.}
    \label{fig:resolution}
\end{figure}

Compared with pixel-domain approaches, the latent-domain design demonstrates better scalability across resolutions. Higher resolutions allow more information to be preserved under the same CBR, resulting in significantly improved performance at $512 \times 512$ compared to $256 \times 256$. As shown in Fig.~\ref{fig:resolution}, a model trained at $256 \times 256$ can still reconstruct high-quality results when applied directly to $512 \times 512$ inputs, even outperforming DiffJSCC trained at $512 \times 512$, highlighting the strong generalization capability of the latent-domain design. Additionally, since the latent representation has a much smaller spatial size than the original image, latent-domain JSCC is substantially more lightweight and efficient than pixel-domain counterparts.

Another key advantage of the proposed framework is its efficiency during the sampling stage. Unlike conventional diffusion models that start from pure Gaussian noise, our method initializes from a semantically meaningful latent representation already close to the true data distribution. Coupled with guidance from the semantic prompt, this approach achieves high-quality reconstructions in as few as $5$ denoising steps—compared to the dozens typically required in standard diffusion models and the 100 steps reported by DiffJSCC. This efficiency makes the proposed method particularly well-suited for wireless communication scenarios, where low latency is critical.

\section{Conclusion}
In this paper, we proposed a diffusion-based semantic communication framework for wireless image transmission. By combining joint source-channel coding with a conditional diffusion model, our approach effectively leverages semantic guidance from multimodal large language models to recover visual information degraded by noisy channels.  Extensive experiments on DIV2K and Kodak datasets under various channel conditions show that our models consistently outperform state-of-the-art JSCC baselines in perceptual quality and distributional fidelity. Notably, the latent-domain design exhibits strong adaptability across resolutions, and the proposed sampling strategy achieves high-quality reconstructions in as few as 5 steps, providing both robustness and low-latency advantages for practical wireless communication scenarios.

\bibliographystyle{IEEEtran}
\bibliography{bib}
\vspace{12pt}
\end{document}